\documentclass[sigconf, screen]{acmart}
\AtBeginDocument{%
  }

\setcopyright{none} 
\renewcommand\footnotetextcopyrightpermission[1]{}
\usepackage{algorithm}
\usepackage{algorithmic}
\usepackage{pifont}

\usepackage{makecell}

\usepackage{array}
\newcolumntype{L}[1]{>{\raggedright\let\newline\\\arraybackslash\hspace{0pt}}m{#1}}
\newcolumntype{C}[1]{>{\centering\let\newline\\\arraybackslash\hspace{0pt}}m{#1}}
\newcolumntype{R}[1]{>{\raggedleft\let\newline\\\arraybackslash\hspace{0pt}}m{#1}}

\usepackage{multirow}

\usepackage{subcaption}

\usepackage{enumitem}
\setlist[itemize]{topsep=2pt, itemsep=2pt, parsep=2pt}

\usepackage{blindtext}

\usepackage{wrapfig}

\DeclareRobustCommand{\tablefont}[1]{#1}
\DeclareRobustCommand{\tablesmall}[1]{\small{#1}}

\usepackage{textcomp}
\newcommand*\whitecircle[1]{\textcircled{\raisebox{-0.0pt}{\small{#1}}}}

\usepackage{svg}
\usepackage{xurl}

\begin{document}

\title{HAWK: Heterogeneity-Aware KV Cache Transfer for Efficient GPU-NPU Phase-Disaggregated LLM Inference}

\author{Junhyeok Park}
\authornote{Work completed while the author was at ETRI; the author is the corresponding author.}
\orcid{0009-0003-3802-7524}
\affiliation{
  \institution{Georgia Institute of Technology}
  \city{Atlanta}  
  \country{USA}
}
\email{junhyeokpark@gatech.edu}

\author{Jungeun Yoon}
\affiliation{
  \institution{Electronics and Telecommunications Research Institute (ETRI)}
  \city{Daejeon}
  \country{Republic of Korea}
}
\email{jeyoon@etri.re.kr}

\author{Kyungmin Kim}
\affiliation{
  \institution{Electronics and Telecommunications Research Institute (ETRI)}
  \city{Daejeon}
  \country{Republic of Korea}
}
\email{kyungmin.kim@etri.re.kr}

\author{Hunsoon Lee}
\affiliation{
  \institution{Electronics and Telecommunications Research Institute (ETRI)}
  \city{Daejeon}
  \country{Republic of Korea}
}
\email{hunsoon@etri.re.kr}

\author{Shinyoung Ahn}
\affiliation{
  \institution{Electronics and Telecommunications Research Institute (ETRI)}
  \city{Daejeon}
  \country{Republic of Korea}
}
\email{syahn@etri.re.kr}

\author{Kangho Kim}
\affiliation{
  \institution{Electronics and Telecommunications Research Institute (ETRI)}
  \city{Daejeon}
  \country{Republic of Korea}
}
\email{khk@etri.re.kr}


\pagestyle{plain}

\begin{abstract}
Large language model (LLM) inference exhibits distinct computational characteristics across its two phases: prefill is compute-intensive, whereas decode is more memory-bound.
As a result, AI accelerators achieve varying levels of efficiency across the two phases.
GPUs are well suited to prefill, but are often underutilized during decode. By contrast, NPUs can utilize their compute resources more effectively during decode.
Motivated by this observation, we explore a heterogeneous architecture that integrates GPUs and NPUs, mapping prefill to GPUs and decode to NPUs to achieve high performance and efficiency.
However, realizing such heterogeneous integration is challenging because LLM phase disaggregation requires KV cache transfer from prefill to decode—that is, from GPUs to NPUs—which is complicated by differences in inter-device communication mechanisms and KV cache tensor layouts.

We propose \textit{HAWK}, a transfer mechanism that enables rapid KV cache transfer across heterogeneous devices. HAWK establishes a remote direct memory access (RDMA) path for direct cross-device transfers while handling inter-device tensor layout discrepancies as part of the transfer process.
In particular, HAWK allocates tensor-layout-aware RDMA buffers on the NPU side to match the target tensor layout. Leveraging these buffers, HAWK orchestrates data transfer from a conventional contiguous GPU RDMA buffer to the corresponding NPU buffers, allowing transferred tensors to be consumed directly without additional data rearrangement while maintaining transfer efficiency.
We implement and evaluate HAWK on a system built with commercially available GPUs and NPUs, achieving performance within 7.88\% of a GPU-only system under the same SLO constraints while delivering up to 1.26$\times$ higher power efficiency and 2.03$\times$ higher cost efficiency.
\end{abstract}


\maketitle

\section{Introduction}

LLMs have become a dominant foundation for modern artificial intelligence systems and are widely used across a broad range of applications, including conversation agents and assistants for coding and research~\cite{model_2025openaigpt5card, model_llama32024herdmodels, model_anthropic2025claude4, model_geminiteam2024gemini15unlockingmultimodal}.
To improve model capabilities, LLMs have grown substantially in scale and complexity, driving a corresponding increase in AI infrastructure requirements~\cite{scaling_kaplan2020scale, scaling_narayanan2021efficientlargescalelanguagemodel, openai2026nextphase, infra_microsoft2025q2, survey_stanford_hai2026aiindex}.
In particular, GPUs have emerged as a primary platform for LLM inference due to their programmability, versatility, and high performance~\cite{nvidia_cuda_guide, nvidia_tensorrtllm, amd_rocm_ai_inference, nvidia_h100_whitepaper, amd_cnda2}.

\begin{table}[!t]
\renewcommand{\arraystretch}{1.00}
  \centering
  \vspace{15pt}
  \resizebox{1.00\columnwidth}{!}{
  \begin{tabular}{C{3.8cm}|C{2.3cm}|C{2.3cm}}
    \hline
    \cline{1-3}
    \multirow{2}{*}{\tablefont{\textbf{Performance}}} & \tablesmall{\textbf{GPU}} & \tablesmall{\textbf{NPU}} \\
    & \tablesmall{NVIDIA A100} &\tablesmall{Tenstorrent p150a} \\
    \hline
    \cline{1-3}
    \tablesmall{Prefill phase throughput \quad\quad\quad (input tokens/$t_{prefill}$)} & \tablefont{\textbf{133.16}} & \tablefont{52.74} \\
    \hline  
    \tablesmall{Decode phase throughput \quad\quad (output tokens/$t_{decode}$)} & \tablefont{\textbf{15.27}} & \tablefont{13.17} \\
    \hline
    \cline{1-3}
    \tablesmall{Decode phase power efficiency \quad (output tokens/$t_{decode}$/kW)} & \tablefont{77.11} & \tablefont{\textbf{107.20}} \\
    \hline
    \tablesmall{Decode phase cost efficiency \quad (output tokens/$t_{decode}$/\$1,000)} & \tablefont{0.55} & \tablefont{\textbf{2.35}} \\
    \hline
    \cline{1-3}
  \end{tabular}
  }
  \vspace{5pt}
    \caption{
    Performance comparison of GPUs and NPUs for LLM inference using requests with 512 input tokens, 128 output tokens, and a batch size of 32.
    The performance gap narrows from 2.52$\times$ in prefill to 1.16$\times$ in decode, while NPUs achieve higher power and cost efficiency during decode.
    Power efficiency is based on device ASIC power; efficiency measurement details are described in Section~\ref{section_eff}.
    }
  \vspace{-15pt}
  \label{table:intro_compare}
\end{table}

\noindent\textbf{GPU Underutilization in LLM Inference: } However, GPU efficiency varies across different phases of LLM inference: prefill and decode.
During prefill, the model processes input tokens in parallel, resulting in general matrix-matrix multiplication (GEMM) operations that efficiently map onto GPU tensor cores~\cite{amd_matrix_core, nvidia_tensor_core_guide, nvidia_h100_whitepaper} and enable high computing utilization.
In contrast, during decode, the model generates one token at a time and the computation is dominated by general matrix-vector multiplication (GEMV). Since GEMV provides limited tile-level parallelism and data reuse, it results in lower compute utilization~\cite{attacc_pim, hybe_npu, systematic_wang2025, pope2023efficiently, dp_hu2024inference, agrawal2024taming, agrawal2023sarathi}.

\noindent\textbf{Opportunities for NPUs:} In this context, neural processing units (NPUs)\footnote{In this paper, NPU broadly refers to specialized AI accelerators other than GPUs.} offer an alternative platform for decode.
NPUs are often designed around dataflow-oriented execution models~\cite{tt_thuning2024attention, npu_sze2017efficient, npu_chen2016eyeriss, tpu_isca17, ascend_hpca, intel_gaudi, aws_inferentia2_arch, avalanche} that provide explicit control over on-chip buffering and compute-memory scheduling. These characteristics can benefit decode by enabling finer-grained coordination and overlap between computation and memory accesses, thereby reducing the impact of off-chip memory accesses during decode.

Table~\ref{table:intro_compare}\footnote{Unless otherwise specified, all experiments use Llama 3.1 70B~\cite{model_llama32024herdmodels}; the rationale for this choice is discussed in Section~\ref{section_software_config}. GPU and NPU measurements each use four devices with a tensor parallelism degree of 4 in vLLM~\cite{vllm}.}
compares LLM inference performance between commodity GPUs (NVIDIA A100~\cite{a100}\footnote{We also evaluate GPUs based on other architectures in Section~\ref{section_h100}.}) and commodity NPUs (Tenstorrent p150a board~\cite{tt_p150a}\footnote{Throughout this paper, we use the Tenstorrent p150a board as the representative NPU platform due to the technical considerations described in Section~\ref{section_hardware_config}. However, HAWK can be applicable to a broader class of NPU architectures.}).
GPUs outperform NPUs during prefill, achieving 2.52$\times$ higher performance. During decode, however, the performance gap narrows to 1.16$\times$.
Notably, NPUs achieve 1.39$\times$ and 4.31$\times$ higher power and cost efficiency, respectively, than GPUs during decode, highlighting their potential for efficient decode serving while maintaining comparable performance.
In particular, modern LLM inference systems increasingly adopt disaggregated prefill-decode architectures~\cite{dp_hu2024inference, distserve, dp_jiang2025hexgen, dp_pdserve, mooncake, dp_ascend_bai2026epd, splitwise_isca, dp_liang2025injecting, dp_hu2024memserve, dp_ruan2025dynaserve, dp_hong2025semi, dp_wu2025arrow, dp_shi2025nexus, dp_wang2025prefill, dp_singh2024efficiently, dp_vendor_chen2025disaggregated}, which decouple the two phases to reduce interference between them and improve system scalability.
In this setting, \textit{a heterogeneous system that uses GPUs for prefill and NPUs for decode can achieve both high performance and efficiency.}

\noindent\textbf{Challenges:} Despite its potential advantages, this heterogeneous system faces several challenges to practical adoption.
\textit{The first challenge} is transferring KV caches generated during GPU prefill to NPUs for decode.
In homogeneous GPU systems, optimized communication software stacks such as NVIDIA NIXL~\cite{nvidia_nixl, nvidia_dynamo} enable direct GPU-to-GPU data transfer via RDMA over InfiniBand or RoCE.
In contrast, heterogeneous data transfer remains relatively underexplored. Moreover, NPUs often rely on diverse, vendor-specific communication mechanisms~\cite{intel_gaudi_white_paper, aws_neuron_rdma, tpu_rdma, tt_architecture_guide}, further complicating interoperability with other devices.

\textit{The second challenge} is handling heterogeneous tensor memory layouts. In LLMs, many data structures, including weights, activations, and KV cache, are represented as tensors~\cite{vllm}.
Accelerators across device types and vendors often employ device-specific tensor memory layouts to reduce memory traffic and optimize computation~\cite{dp_vendor_chen2025disaggregated, tt_tensor_layout, xla_tpu_layout, ascend_layout, qualcomm_tensor_layout}.
Resolving such tensor layout discrepancies requires additional data rearrangement, complicating data transfer and increasing latency.

\noindent\textbf{Enabling Heterogeneous LLM Inference:}
This paper presents \textit{HAWK} (\underline{H}eterogeneity-\underline{AW}are \underline{K}ey-Value Transfer), a mechanism that enables high-performance LLM inference across heterogeneous devices.
HAWK establishes a direct data transfer path between GPU and NPU memory by connecting both to a common RDMA stack.
To handle tensor layout discrepancies, HAWK applies a transfer-aware transformation to the GPU KV cache, partially aligning it with the target NPU layout while grouping data destined for the same NPU DRAM channel into contiguous portions for efficient RDMA transfer. On the NPU side, HAWK allocates one RDMA buffer per DRAM channel according to the target tensor layout. Transferring each contiguous portion to its corresponding NPU buffer completes the remaining layout alignment, enabling decode without post-transfer layout conversion.

In our evaluation, at an aggregated transfer size of 2560 MB, HAWK achieves a 25.40$\times$ speedup over naive host-staged GPU-to-NPU transfers, with end-to-end latency differing by only 0.39\% from GPU-to-GPU RDMA.
Compared with a GPU-only system, our heterogeneous system with HAWK achieves up to 1.26$\times$ higher power efficiency and 2.03$\times$ higher cost efficiency, with a performance difference of at most 7.88\% under the same SLO constraints.
To the best of our knowledge, this work is the first to identify and address practical challenges in enabling direct GPU-to-NPU KV cache transfer for disaggregated LLM serving on a real system.
While this work focuses on KV cache transfer, HAWK can also support other tensor objects, such as weights and activations, required by other forms of disaggregated LLM serving.

Overall, this paper makes the following contributions.

\begin{itemize}[leftmargin=*]
\item
We characterize the phase-dependent execution behavior of GPUs and NPUs for LLM inference. GPUs achieve high compute utilization during prefill but are substantially underutilized during decode.
In contrast, NPUs exhibit higher compute utilization while achieving decode performance comparable to GPUs, providing a more efficient design point consistent with their dataflow-oriented architectures.

\item
We propose a heterogeneous GPU–NPU serving system that assigns prefill to GPUs and decode to NPUs to achieve high performance and efficiency. We identify cross-device data transfer as a critical bottleneck in this design, motivating the need for a high-performance heterogeneous transfer mechanism.

\item
We present \textit{HAWK}, a heterogeneity-aware KV cache transfer mechanism. HAWK leverages an RDMA-based transfer path that bypasses host staging in the data path and handles cross-device tensor-layout mismatches through transfer-aware tensor transformation and tensor-layout-aware RDMA buffer allocation.
\end{itemize}


\section{Background}

\subsection{LLM Inference}

\begin{figure}[!t]
\centering
\scalebox{1.00}{\includegraphics[width=1\columnwidth]{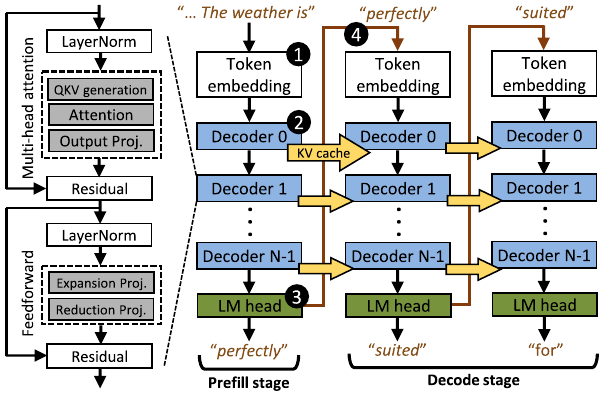}}
\vspace{-8pt}
\caption{An overview of the LLM inference process.}
\label{llm_overview}
\vspace{-10pt}
\end{figure}

Figure~\ref{llm_overview} illustrates the inference process of a conventional decoder-only Transformer, which consists of two phases: prefill and decode. Both phases follow the same sequence of a token embedding layer, multiple decoder layers, and a language modeling (LM) head, but differ in the number of tokens processed.
During the prefill phase, all input tokens are first converted into embedding vectors by the token embedding layer (\ding{182}). The resulting embeddings are then processed by multiple decoder layers, each consisting of layer normalization, multi-head attention, residual connections, and a feedforward network (FFN) (\ding{183}), to capture inter-token relationships, incorporate contextual information, and refine token representations. The output of the decoder layers is then passed through the LM head to produce token scores, from which the next token is selected (\ding{184}). The selected token is subsequently fed back into the token embedding layer as input, initiating the autoregressive decode phase (\ding{185}), which continues until an end-of-sequence (EOS) token is generated or the maximum generation length is reached.

In particular, multi-head attention within each decoder layer computes Query, Key, and Value (QKV) activations from the input, which serve as the core intermediate representations for attention computation and inter-token interaction~\cite{attention2017need}. The Key and Value activations generated for each token remain unchanged during subsequent decoding steps and can therefore be reused. To exploit this property, LLM inference stores these activations in memory as a key-value (KV) cache and retrieves them in later decoding steps, avoiding redundant computation. The KV cache grows incrementally as new tokens are generated, with the Key and Value activations of each new token appended to the existing cache.

\begin{table}[!t]
\renewcommand{\arraystretch}{1.00}
  \centering
  \resizebox{1.00\columnwidth}{!}{
  \begin{tabular}{C{2.3cm}|C{2.8cm}|C{2.2cm}}
    \hline
    \cline{1-3} 
    \tablesmall{\textbf{Operation}} & \tablesmall{\textbf{Operands}} & \tablesmall{\textbf{Output}} \\
    \hline
    \cline{1-3} 
    \tablesmall{QKV projection} & \tablesmall{[\textit{B}, \textit{L}, \textit{H}], [\textit{H}, \textit{3H}]} & \tablesmall{[\textit{B}, \textit{L}, \textit{3H}]} \\
    \hline
    \multirow{2}{*}{\tablesmall{Attention}} & \tablesmall{[\textit{B}, \textit{L}, \textit{H}], [\textit{B}, \textit{H}, \textit{$L_t$}]} & \tablesmall{[\textit{B}, \textit{L}, \textit{$L_t$}]} \\
    & \tablesmall{[\textit{B}, \textit{L}, \textit{$L_t$}], [\textit{B}, \textit{$L_t$}, \textit{H}]} &\tablesmall{[\textit{B}, \textit{L}, \textit{H}]} \\
    \hline 
    \tablesmall{Output projection} & \tablesmall{[\textit{B}, \textit{L}, \textit{H}], [\textit{H}, \textit{H}]} & \tablesmall{[\textit{B}, \textit{L}, \textit{H}]} \\
    \hline
    \multirow{2}{*}{\tablesmall{FFN projections}} & \tablesmall{[\textit{B}, \textit{L}, \textit{H}], [\textit{H}, \textit{F}]} & \tablesmall{[\textit{B}, \textit{L}, \textit{F}]} \\
    & \tablesmall{[\textit{B}, \textit{L}, \textit{F}], [\textit{F}, \textit{H}]} &\tablesmall{[\textit{B}, \textit{L}, \textit{H}]} \\
    \hline 
    \cline{1-3} 
  \end{tabular}
  }
  \vspace{5pt}
    \caption{
    Shape of the operands and output tensors for each operation in a decoder layer, assuming single-head attention for simplicity. Here, \textit{$B$}, \textit{$L$}, \textit{$L_t$}, \textit{$H$}, and \textit{$F$} represent the batch size, input token length, total token length, model hidden dimension, and FFN hidden dimension, respectively. The operand shapes depend strongly on $L$, determining whether the resulting computation is more GEMM-like or GEMV-like.
    }
  \vspace{-15pt}
  \label{table:operations}
\end{table}

\subsection{Characteristics of Prefill and Decode}

The dominant computations in the decoder layer, including QKV projections, attention computation, output projection, and FFN projections, depend on the input shape~\cite{agrawal2023sarathi, attacc_pim}, as shown in Table~\ref{table:operations}. During prefill, all input tokens are processed in parallel ($L$ equals the initial input length), producing a matrix of embedding vectors; consequently, these computations are dominated by GEMM.
During decode, in contrast, the model processes one newly generated token at a time ($L$ equals 1), producing a vector; accordingly, the dominant computations are mapped to GEMV, which exhibits lower arithmetic intensity and reduced data reuse relative to GEMM.
Moreover, decode requires fetching KV caches from previously processed tokens, which introduces more memory traffic than prefill.

\begin{figure}[!t]
\begin{subfigure}{0.58\columnwidth}
    \includegraphics[width=\linewidth]{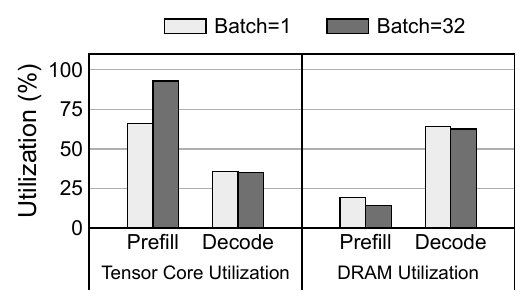}
\caption{}
    \label{motiv_gpu_kernel}
\end{subfigure}
\hspace{1pt}
\begin{subfigure}{0.39\columnwidth}
    \includegraphics[width=\linewidth]{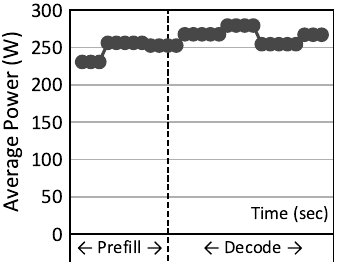}
\caption{}
    \label{motiv_gpu_power}
\end{subfigure}
\vspace{-10pt}
    \caption{
    (a) GPU tensor core and DRAM utilization of the QKV projection kernel and (b) GPU board-level power consumption during prefill and decode, using requests with 512 input tokens and 128 output tokens.
    GPUs achieve high computational efficiency during prefill but lower efficiency during decode, while power consumption does not closely track computational efficiency across phases.
    }
\label{motive_gpu_profile}
\vspace{-7pt}
\end{figure}

\begin{table}[!t]
\renewcommand{\arraystretch}{0.95}
  \centering
  \resizebox{0.98\columnwidth}{!}{
  \begin{tabular}{C{2.3cm}|C{2.5cm}C{2.5cm}}
    \hline
    \cline{1-3}
    \multirow{2}{*}{\tablefont{\textbf{Operations}}} & \multicolumn{2}{c}{\tablesmall{\textbf{GPU Utilization (\%)}}} \\
    \cline{2-3}
    & \tablesmall{\textbf{Tensor Core}} & \tablesmall{\textbf{DRAM}} \\
    \hline
    \cline{1-3}
    \tablesmall{QKV projection} & \tablefont{35.19} & \tablefont{62.48} \\
    \hline
    \tablesmall{Attention}& \tablefont{22.22} & \tablefont{36.65} \\
    \hline  
    \tablesmall{Output projection} & \tablefont{33.30} & \tablefont{59.85} \\
    \hline
    \tablesmall{FFN projections} & \tablefont{38.77} & \tablefont{72.40} \\
    \hline
    \cline{1-3}
  \end{tabular}
  }
  \vspace{3pt}
\caption{
GPU tensor core and DRAM utilization of kernels in a decoder layer during decode, using requests with 512 input tokens, 128 output tokens, and a batch size of 32. All kernels exhibit low tensor core utilization due to the limited parallelism during decode.
}
  \vspace{-15pt}
  \label{table:gpu_kernels}
\end{table}

\section{Motivation}

\subsection{Phase-Dependent GPU Efficiency}

GPUs are a primary platform for LLM inference due to their high performance and mature software ecosystem~\cite{vllm, sglang, nvidia_tensorrtllm, infinigen}. However, GPU computational efficiency varies across inference phases due to the distinct characteristics of prefill and decode.
We profile GPU kernels in a decoder layer during prefill and decode on A100 GPUs using Nsight Compute~\cite{nsight_compute}.
Figure~\ref{motiv_gpu_kernel} shows tensor core and DRAM utilization for the QKV projection kernel as a representative example.
Across batch sizes 1 and 32, the kernel achieves high tensor core utilization (up to 92.92\%) with relatively low DRAM utilization (up to 19.26\%) during prefill, consistent with GEMM-dominated computation that maps efficiently onto GPU tensor cores. During decode, in contrast, the kernel exhibits lower tensor core utilization (up to 35.90\%) and higher DRAM utilization (up to 64.09\%), consistent with GEMV-dominated computation characterized by low arithmetic intensity and limited data reuse.
Table~\ref{table:gpu_kernels} shows that other kernels in a decoder layer also exhibit low computational efficiency during decode for similar reasons.

Although request batching can make GEMV execution more GEMM-like by increasing $B$ in Table~\ref{table:operations}, these benefits emerge only at sufficiently large batch sizes~\cite{attacc_pim}, as the performance gains from small-batch execution are limited, as shown in Figure~\ref{motiv_gpu_kernel}.
In practice, however, such large batch sizes are often infeasible due to memory capacity constraints: the KV cache size in elements, $2 \times L_t \times N_l \times N_h \times N_{hd} \times B$, grows linearly with batch size, where $N_l$, $N_h$, and $N_{hd}$ denote the number of decoder layers, the number of KV heads, and the head dimension, respectively.

Furthermore, GPU power consumption remains relatively stable across the two phases, as shown in Figure~\ref{motiv_gpu_power}.
Because tensor cores represent a significant component of GPU power consumption when active~\cite{accelwatch}, the lower tensor core utilization during decode might be expected to reduce power consumption. However, the power consumption does not show a corresponding decrease when transitioning from prefill to decode.
Consequently, decode on GPUs exhibits not only lower computational efficiency but also lower power efficiency than prefill.

\begin{table}[!t]
\renewcommand{\arraystretch}{0.95}
\centering
\resizebox{0.98\columnwidth}{!}{
\begin{tabular}{C{2.3cm}|C{1.27cm}C{1.27cm}|C{1.23cm}C{1.23cm}}
\hline
\cline{1-5}
\multirow{4}{*}{\tablefont{\textbf{Operations}}}
& \multicolumn{2}{c|}{%
  \multirow{2}{*}{\shortstack{
    \tablesmall{\textbf{Compute Utilization}} \\
    \tablesmall{(\% of peak throughput)}
  }}}
& \multicolumn{2}{c}{%
  \multirow{2}{*}{\shortstack{
    \tablesmall{\textbf{Power Efficiency}} \\
    \tablesmall{(TFLOPS/kW)}
  }}} \\
& \multicolumn{2}{c|}{}
& \multicolumn{2}{c}{} \\
\cline{2-5}

& \multirow{2}{*}{\shortstack{
    \tablesmall{\textbf{GPU}} \\ \tablesmall{A100}}}
& \multirow{2}{*}{\shortstack{
    \tablesmall{\textbf{NPU}} \\ \tablesmall{p150a}}}
& \multirow{2}{*}{\shortstack{
    \tablesmall{\textbf{GPU}} \\ \tablesmall{A100}}}
& \multirow{2}{*}{\shortstack{
    \tablesmall{\textbf{NPU}} \\ \tablesmall{p150a}}} \\
& & & & \\
\hline
\cline{1-5}

\tablesmall{QKV projection}
& \tablefont{11.93} & \tablefont{79.06}
& \tablefont{226.82} & \tablefont{167.59} \\
\hline

\tablesmall{Attention}
& \tablefont{1.67} & \tablefont{8.34}
& \tablefont{22.64} & \tablefont{28.34} \\
\hline

\tablesmall{Output projection}
& \tablefont{11.20} & \tablefont{74.33}
& \tablefont{213.00} & \tablefont{157.55} \\
\hline

\tablesmall{FFN projections}
& \tablefont{13.44} & \tablefont{81.46}
& \tablefont{255.47} & \tablefont{258.99} \\
\hline
\cline{1-5}
\end{tabular}
}
  \vspace{3pt}
\caption{
Comparison of compute utilization and power efficiency across
kernels in a decoder layer during decode on GPUs and NPUs,
using requests with 512 input tokens, 128 output tokens,
and a batch size of 32.
NPUs achieve higher compute utilization across all the
kernels through phase-aware core allocation and higher power efficiency for attention and FFN
projections, which account for a substantial fraction of
LLM decoding time~\cite{systematic_wang2025}.
Power efficiency is based on device ASIC power; measurement details are provided in Section~\ref{section_eff}.
}
  \vspace{-5pt}
  \label{table:kernel_compare}
\end{table}

\subsection{Potential of NPUs for Efficient Decode}

NPUs are specialized accelerators designed for AI workloads. While their detailed designs vary, most broadly adopt dataflow-oriented architectures that explicitly manage computation and data movement to maximize data reuse and reduce the impact of off-chip memory accesses~\cite{tpu_isca17, ascend_hpca, intel_gaudi, aws_inferentia2_arch}. These characteristics can be particularly beneficial for GEMV-like computation during decode.

Table~\ref{table:kernel_compare} compares the compute utilization and power efficiency of GPUs (A100) and NPUs (p150a) for representative kernels in a decoder layer during decode.
Compute utilization is calculated by dividing the achieved throughput (TFLOPS) by the peak throughput of the compute cores used by each kernel and multiplying by 100.
During decode, the evaluated NPUs use fewer compute cores than during prefill (e.g., 8 versus 64 for FFN kernels~\cite{tt_metal}) and select cores closer to DRAM to better accommodate the limited parallelism and memory-intensive nature of GEMV-like computation.
By concentrating computation on fewer cores, they maintain high utilization of the assigned cores, exceeding GPU utilization across all evaluated kernels.
While using fewer cores may limit kernel throughput, it can also improve power efficiency by reducing the number of active compute cores.
In our evaluation, the NPU achieves higher power efficiency for attention and FFN kernels, which account for a substantial fraction of LLM decoding time~\cite{systematic_wang2025}.
Although we evaluate one NPU platform, other NPU architectures may achieve similar efficiency gains through different mechanisms.
When these efficiency gains are achieved while maintaining comparable performance, as shown in Table~\ref{table:intro_compare}, NPUs can be considered an attractive option for efficient decode serving.

\begin{figure}[!t]
\centering
\vspace{-2pt}
\scalebox{1.00}{\includegraphics[width=1\columnwidth]{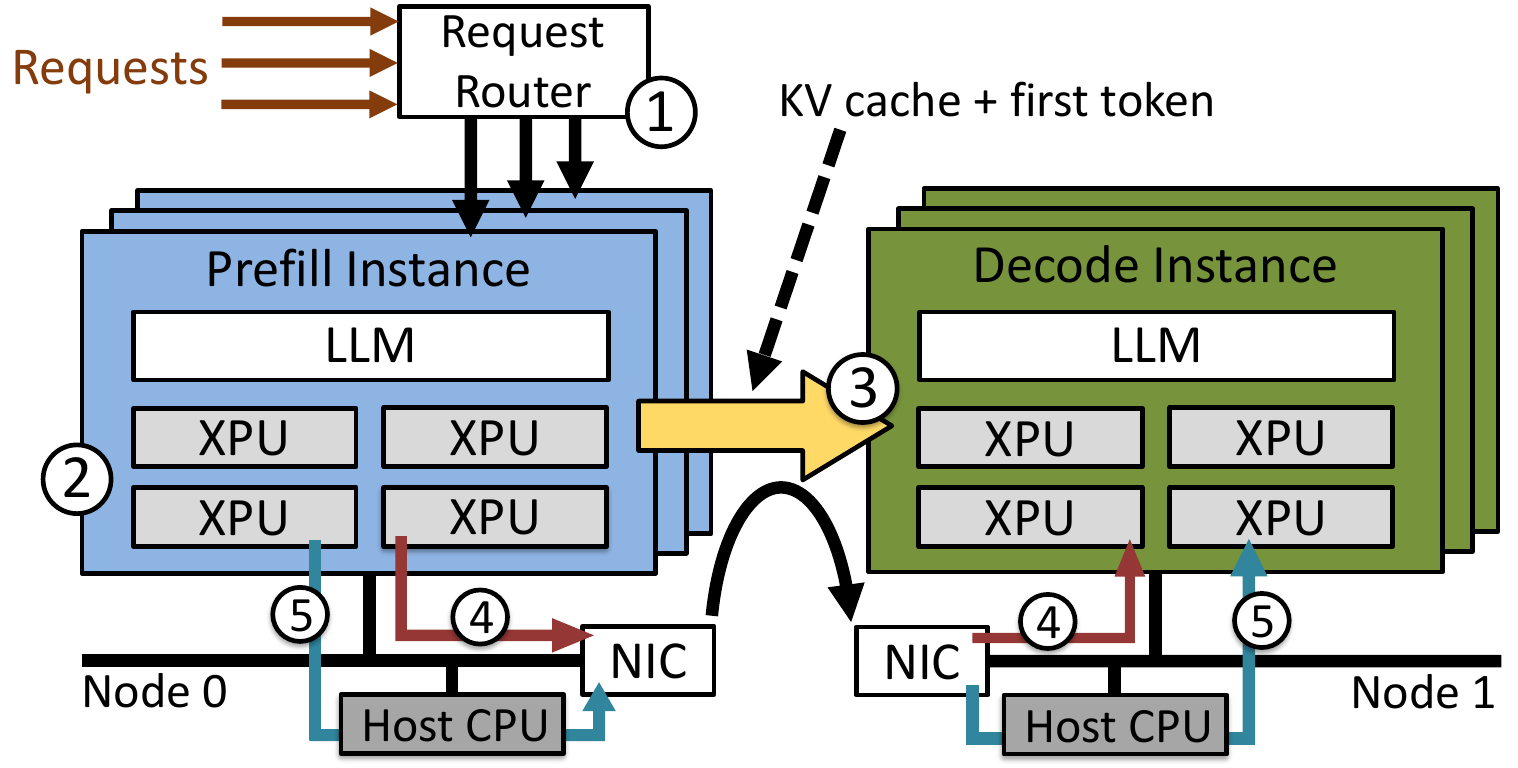}}
\vspace{-15pt}
\caption{An overview of a prefill-decode phase-disaggregated LLM serving system.}
\label{disaggregation}
\vspace{-4pt}
\end{figure}

\subsection{LLM Disaggregation with GPUs and NPUs}

Figure~\ref{disaggregation} illustrates a phase-disaggregated LLM serving system that separates the prefill and decode phases on different computing resources~\cite{dp_hu2024inference, distserve, dp_jiang2025hexgen, dp_pdserve, mooncake, dp_ascend_bai2026epd, splitwise_isca, dp_liang2025injecting, dp_hu2024memserve, dp_ruan2025dynaserve, dp_hong2025semi, dp_wu2025arrow, dp_shi2025nexus, dp_wang2025prefill, dp_singh2024efficiently, dp_vendor_chen2025disaggregated}.
An LLM replica is loaded onto each computing resource, typically on separate server nodes, and the request router forwards incoming requests to the prefill instance (\ding{172}). The prefill instance generates the first token and KV cache for the input tokens (\ding{173}), which are then transferred to the decode instance for subsequent token generation (\ding{174}).
By isolating the two phases, this system reduces inter-phase interference, thereby stabilizing tail latency and improving request goodput~\cite{distserve, real_dp_cloudflare, mooncake}. It also enables flexible scaling of prefill and decode instances and phase-specific resource allocation, making it attractive for production LLM serving~\cite{splitwise_isca, mooncake}.
As a result, many LLM serving software stacks actively support phase disaggregation~\cite{vllm, sglang, nvidia_dynamo, nvidia_tensorrtllm, llmd, aws_dp}, which has also been widely adopted in real-world LLM serving platforms~\cite{mooncake, real_dp_cloudflare, real_dp_together, llmd}.

Leveraging this approach, we explore a heterogeneous LLM serving system that maps prefill to GPUs and decode to NPUs according to their computational characteristics. GPUs provide high performance and efficiency during prefill, while NPUs offer comparable decode performance with higher efficiency, enabling the system to achieve both high performance and efficiency.

\subsection{Challenges: Handling Heterogeneity}

Despite the potential benefits of a heterogeneous GPU–NPU system, several challenges hinder its practical deployment.
\textit{The first challenge} is cross-device data transfer, particularly for KV caches. In a GPU-only system, GPUs support RDMA-based transfers~\cite{nvidia_gpudirect_rdma} through optimized communication software~\cite{nvidia_nixl, mooncake}, allowing the network interface card (NIC) to directly access GPU memory, as shown in Figure~\ref{disaggregation} (\ding{175}).
However, in a heterogeneous system, GPUs and NPUs rely on distinct software stacks and communication mechanisms~\cite{intel_gaudi_white_paper, aws_neuron_rdma, tpu_rdma, tt_architecture_guide}, which may not directly interoperate.
A straightforward solution is to use the host as a staging point, which avoids the need for direct interoperability (\ding{176}). However, this approach incurs substantial latency from additional copies between the host and each device, as well as host intervention.
We observed that, even between GPUs, host-staged transfer incurs 1.83$\times$ to 5.25$\times$ the latency of direct transfer as the transfer size increases from 20 MB to 2.5 GB, adding substantial overhead to LLM serving.

\begin{figure}[!t]
\begin{subfigure}{0.48\columnwidth}
    \includegraphics[width=\linewidth]{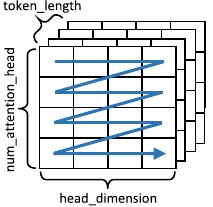}
\caption{GPU KV cache tensor layout.}
    \label{gpu_tensor}
\end{subfigure}
\hspace{4pt}
\begin{subfigure}{0.48\columnwidth}
    \includegraphics[width=\linewidth]{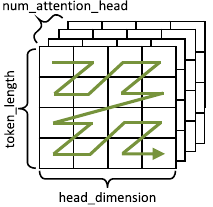}
\caption{NPU KV cache tensor layout.}
    \label{npu_tensor}
\end{subfigure}
\vspace{-17pt}
\caption{
Comparison of tensor layouts across different devices. The arrows represent tensor elements in contiguous virtual addresses.
Each device uses its own ordering of tensor dimensions and virtual-address mapping of tensor elements.
}
\vspace{-5pt}
\label{tensor_layout}
\end{figure}

\textit{The second challenge} is handling differences in KV cache tensor layouts across devices. A tensor layout defines the ordering and shape of tensor dimensions, as well as how tensor elements are mapped to the virtual address space and distributed across physical memory channels. Different accelerators employ device-specific tensor layouts to optimize data movement and computation, and these differences can be particularly pronounced between GPUs and NPUs~\cite{tt_tensor_layout, xla_tpu_layout, ascend_layout, qualcomm_tensor_layout} (e.g., row- or column-major layouts on GPUs versus tiled layouts on NPUs), as illustrated in Figure~\ref{tensor_layout}.
Consequently, GPU-to-NPU tensor transfers require additional layout transformation to match the target NPU layout, introducing extra latency.
One way to avoid this overhead is to force one device to adopt the tensor layout of the other. However, because each tensor layout is optimized for its device, enforcing a unified layout can reduce memory access efficiency and requires modifications to existing compute kernels, potentially degrading their performance~\cite{vattention,conserve}.

\begin{figure}[!t]
\centering
\scalebox{1.00}{\includegraphics[width=1\columnwidth]{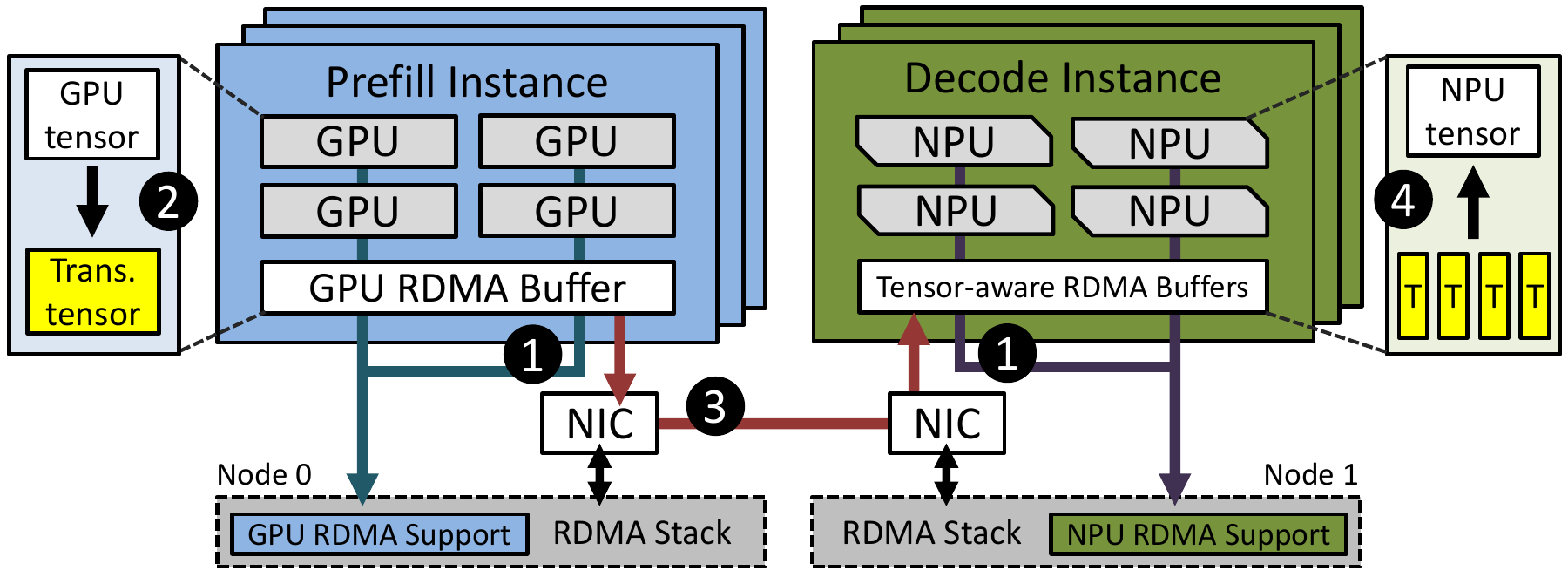}}
\vspace{-10pt}
\caption{An overview of a GPU-NPU phase-disaggregated LLM inference system with HAWK.}
\label{main_overview}
\end{figure}

\section{Heterogeneity-Aware KV Cache Transfer}

\textbf{Our goal} is to enable an efficient GPU–NPU heterogeneous LLM inference system. To this end, we present \textit{HAWK}, a heterogeneity-aware KV cache transfer mechanism that enables direct, high-performance tensor transfer between GPUs and NPUs. 

Figure~\ref{main_overview} illustrates an overview of a heterogeneous system with HAWK.
HAWK builds on the Linux RDMA stack and establishes a common RDMA path for GPU and NPU memory by integrating device-specific RDMA memory support (\ding{182}), enabling the NIC on each node to directly access device memory without host intervention.
Once the GPU prefill instance generates the KV cache, HAWK transforms the GPU KV tensors into transfer-friendly aligned tensors~(\ding{183}). The transformation partially aligns tensor elements with the target NPU layout while maximizing the size of each RDMA transfer. 
HAWK then transfers the transformed tensors from a conventional GPU RDMA buffer into multiple layout-aware RDMA buffers on the NPU, which are organized according to the target tensor layout~(\ding{184}). 
This organization allows the partially aligned tensors to naturally conform to the target NPU layout during transfer, enabling the received tensors to be consumed directly during decode without additional layout transformation~(\ding{185}).

\begin{figure}[!t]
\centering
\scalebox{1.00}{\includegraphics[width=1\columnwidth]{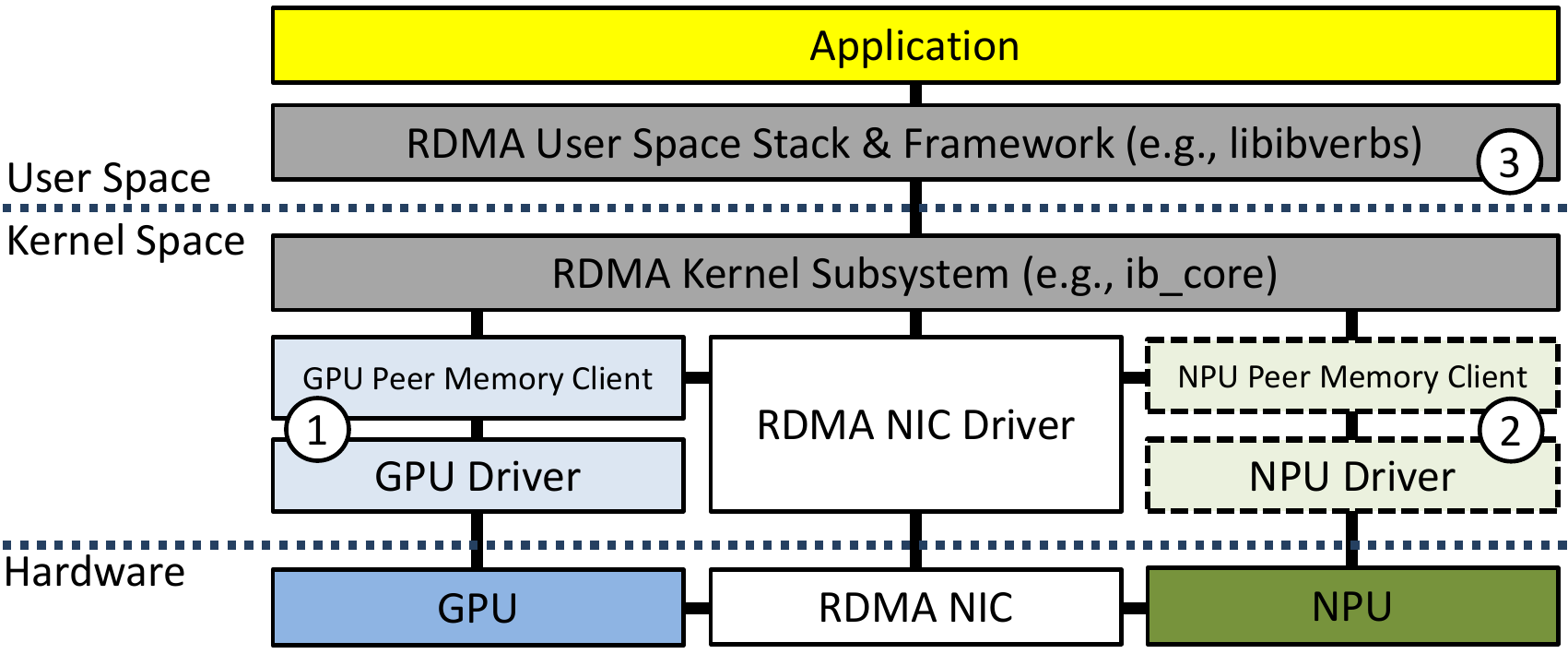}}
\vspace{-10pt}
\caption{RDMA stack providing a common transfer path between GPU and NPU device memory through device-specific peer memory support.
Dotted-outline blocks indicate components extended as part of HAWK.
}
\label{rdma_stack}
\vspace{-4pt}
\end{figure}

\begin{table}[!t]
\renewcommand{\arraystretch}{1.00}
  \centering
  \resizebox{1.0\columnwidth}{!}{
  \begin{tabular}{R{2.2cm}|L{5.9cm}}
    \hline
    \cline{1-2}
    \tablefont{\textbf{Callbacks}} & \tablefont{\textbf{Description}} \\
    \hline  
    \cline{1-2}
    \tablesmall{\texttt{npu\_acquire}} & \tablesmall{Check whether the address range requested by the RDMA subsystem belongs to NPU device memory} \\
    \hline
    \tablesmall{\texttt{npu\_get\_pages}} & \tablesmall{Retrieve the NPU memory pages for the address range requested by the RDMA subsystem} \\
    \hline
    \tablesmall{\texttt{npu\_dma\_map}} & \tablesmall{Map the retrieved NPU memory pages to DMA addresses accessible to the NIC} \\
    \hline
    \tablesmall{\texttt{npu\_dma\_unmap}} & \tablesmall{Release DMA mappings for the NPU memory pages} \\
    \hline
    \tablesmall{\texttt{npu\_put\_pages}} & \tablesmall{Release the NPU memory pages retrieved by \texttt{npu\_get\_pages}} \\
    \hline
    \cline{1-2}
  \end{tabular}
  }
  \vspace{5pt}
\caption{
Major callbacks in the NPU peer memory client for exposing NPU memory mappings to the RDMA subsystem.
}
  \vspace{-15pt}
  \label{table:npu_peer_client}
\end{table}

\subsection{Building a Direct Data Path Across Devices}

The essential step toward efficient data transfer is to establish a high-performance transfer path between computing devices.
To this end, HAWK connects GPU and NPU device memory through a common Linux RDMA stack spanning kernel space (e.g., \texttt{ib\_core}~\cite{ib_core}) and user space (e.g., \texttt{libibverbs}~\cite{libibverbs_library_ibm}), as shown in Figure~\ref{rdma_stack}.
On the GPU side, HAWK leverages the GPU peer memory client~\cite{nvidia_gpudirect_rdma, nvidia_peermem} (\ding{172}), whereas on the NPU side, it extends the NPU driver with a corresponding peer memory client (\ding{173}).
Table~\ref{table:npu_peer_client} summarizes the implementation of the NPU-side RDMA stack.
These clients expose device memory to the RDMA subsystem by providing the memory-mapping information required for memory registration, enabling the NIC to directly access the registered memory.
Building on this kernel-level support and a common RDMA stack, various transfer engines~\cite{mooncake, mao2025uccl, zhou2025uccl_extensible} abstract low-level RDMA primitives and orchestrate runtime data movement (\ding{174}). Together, these components establish a high-performance, flexible data transfer path across heterogeneous devices.

\begin{figure}[!t]
\centering
\scalebox{1.00}{\includegraphics[width=1\columnwidth]{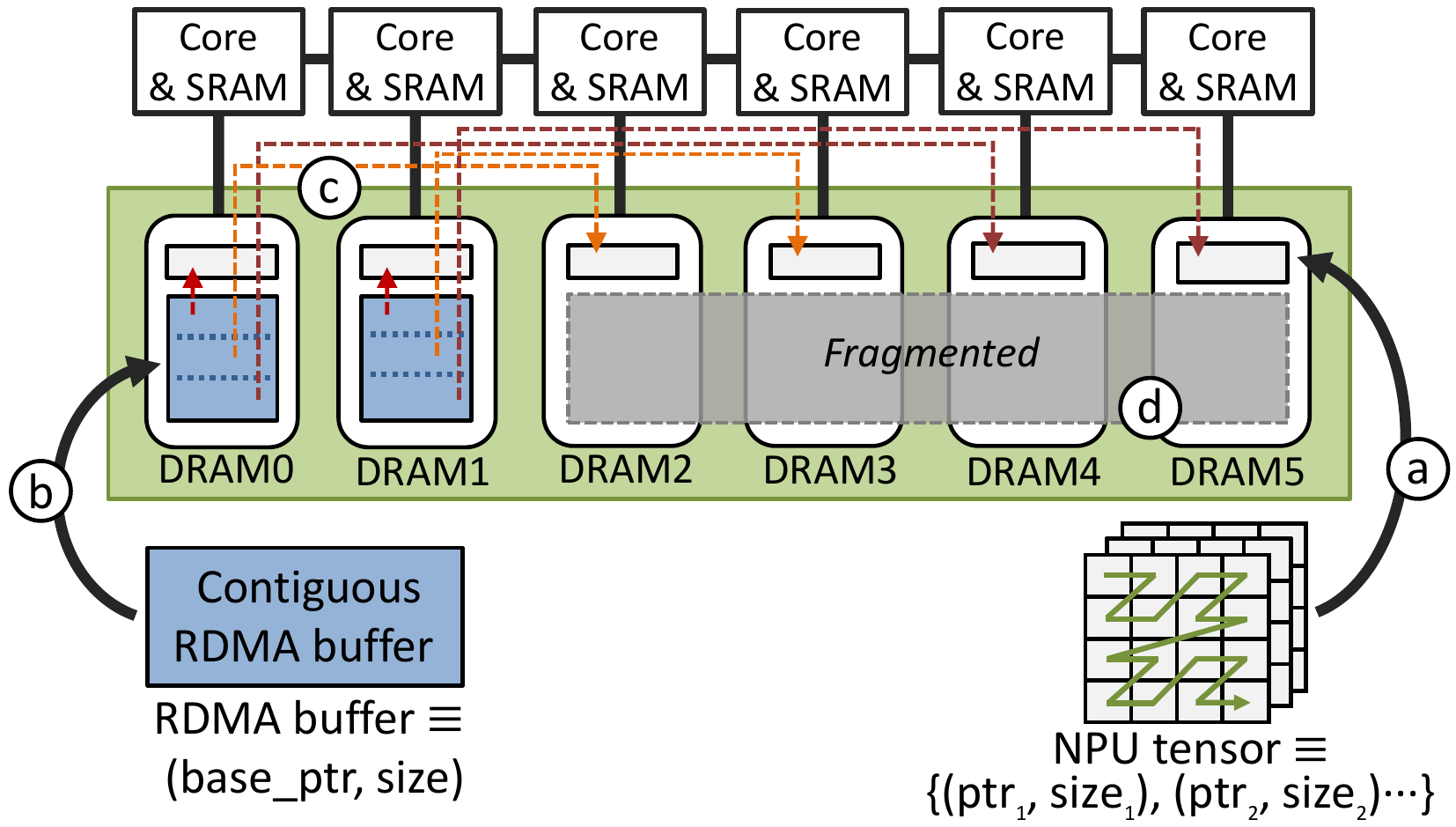}}
\vspace{-17pt}
\caption{
Physical memory layouts of an NPU tensor and an equally sized contiguous buffer.
An NPU tensor is distributed across multiple DRAM channels, whereas a contiguous buffer follows a different channel layout.
}
\label{npu_layout}
\vspace{-5pt}
\end{figure}
\begin{table}[!t]
\renewcommand{\arraystretch}{0.95}
  \centering
  \resizebox{1.0\columnwidth}{!}{
  \begin{tabular}{R{3.3cm}|L{4.8cm}}
    \hline
    \cline{1-2}
    \tablefont{\textbf{APIs}} & \tablefont{\textbf{Description}} \\
    \hline  
    \cline{1-2}
    \tablesmall{\texttt{map\_buffer\_dma}} & \tablesmall{Create DMA mappings for a tensor-like device buffer} \\
    \hline
    \tablesmall{\texttt{get\_device\_num\_bank}} & \tablesmall{Get the number of DRAM banks (channels) in an NPU device} \\
    \hline
    \tablesmall{\texttt{get\_buffer\_address\_bank}} & \tablesmall{Get the buffer base address allocated in a specific DRAM bank (channel)} \\
    \hline
    \tablesmall{\texttt{get\_buffer\_size\_bank}} & \tablesmall{Get the buffer size allocated in a specific DRAM bank (channel)} \\
    \hline
    \tablesmall{\texttt{unmap\_buffer\_dma}} & \tablesmall{Release DMA mappings for a tensor-like device buffer} \\
    \hline
    \cline{1-2}
  \end{tabular}
  }
  \vspace{5pt}
  \caption{Major C++ APIs for supporting tensor-layout-aware RDMA buffer allocation.}
  \vspace{-20pt}
  \label{table:tensor_aware}
\end{table}

\subsection{Tensor-Layout-Aware Buffer Allocation}
\label{section_buffer}

RDMA buffers are memory regions used to store data for RDMA operations.
In conventional RDMA use cases, allocating a single large virtually contiguous buffer is practical because it simplifies buffer metadata management and RDMA request construction compared with using multiple scattered buffers.
However, this design can be less efficient for NPUs, especially when transferring tensors.
For GPUs, tensors are distributed across multiple DRAM channels at a fixed granularity (e.g., 256 B) to increase memory-level parallelism~\cite{leveraging_park25, get_vaelly_2018isca} while remaining mapped to virtually contiguous address ranges, making a single RDMA buffer well suited to the GPU memory organization.
NPUs also distribute tensors across multiple DRAM channels, but explicitly control which tensor chunks are placed in each partition rather than relying on a fixed-size mapping, improving memory locality and reducing traffic between off-chip memory and on-chip SRAM.
Therefore, an NPU tensor has a distinct memory layout and address mapping compared with a conventional virtually contiguous buffer.

Figure~\ref{npu_layout} illustrates the physical memory layouts of an NPU tensor and an equally sized contiguous buffer. While the tensor is distributed across multiple DRAM channels, its address is represented by a set of pointer-size pairs, each referring to a chunk placed in a different DRAM channel~(\whitecircle{a}), as expressed in Eq~(\ref{equation_npu_tensor}).
\begin{equation}
\begin{aligned}
\texttt{npu\_tensor* T}
&\equiv
\{(ptr_{ch0}, size_{ch0}),
  (ptr_{ch1}, size_{ch1}),
  \ldots\}.
\end{aligned}
\label{equation_npu_tensor}
\end{equation}
In contrast, a virtually contiguous buffer is represented by a single
pointer-size pair, as expressed in Eq~(\ref{equation_npu_buffer}),
and follows a linear address layout that is not naturally aligned
with the NPU tensor layout~(\whitecircle{b}).
\begin{equation}
\texttt{buffer* B}
\equiv
(ptr, size).
\label{equation_npu_buffer}
\end{equation}
These discrepancies in address mapping and memory organization between GPUs and NPUs introduce two major issues.
\textit{The first issue} is that, even if a GPU tensor is transformed to match the target NPU tensor layout in dimensions and element ordering, transferring it through a conventional RDMA buffer may still require additional post-transfer adjustment before it can be used as an NPU tensor object~(\whitecircle{c}).
\textit{The second issue} is that using a conventional virtually contiguous RDMA buffer on an NPU can cause severe memory fragmentation. Because the buffer layout does not follow the NPU tensor layout, memory usage can become imbalanced across DRAM channels, leaving some channels exhausted while capacity remains available in others. This imbalance can prevent subsequent tensor allocations~(\whitecircle{d}). Even if new tensors are allocated only from partitions with sufficient remaining capacity, such selective allocation can disrupt the intended data placement and reduce data-movement efficiency, ultimately degrading performance.

To address this issue, HAWK allocates NPU RDMA buffers according to the target NPU tensor layout, using one virtually contiguous buffer per DRAM channel with corresponding software support, as shown in Table~\ref{table:tensor_aware}. This layout-aware allocation allows transferred data to be consumed directly from the RDMA buffers or converted into the target tensor with only limited intra-channel data movement, avoiding costly data rearrangement across multiple DRAM channels.
Although using multiple virtually non-contiguous buffers can increase buffer-management overhead and the number of RDMA requests required to transfer a tensor, the number of buffers is bounded by the number of DRAM channels in the NPU. This count is typically around ten or fewer, and each per-channel buffer remains contiguous. As a result, the additional buffer-management and RDMA-request overhead remains modest relative to a conventional single-buffer design.

\subsection{Transfer-Aware Tensor Transformation}

GPU tensors must be transformed to match the target NPU tensor layout for correct execution. HAWK decomposes this transformation into two steps to improve transfer efficiency by leveraging tensor-layout-aware RDMA buffers on the NPU side. In the first step, HAWK partially aligns the GPU tensor with the target NPU layout by reordering the tensor dimensions and elements while arranging portions destined for the same per-channel NPU buffer contiguously~(\ding{182}), as expressed in Eq~(\ref{equation_transform_1}). 
Here, $-1$ denotes an inferred dimension, and
$\texttt{tile\_chunk}$ is the NPU tile size in elements.
\begin{equation}
\begin{aligned}
\texttt{gpu\_tensor* G}:
\mathrm{shape}(\texttt{G})
&=
[N_l, L, N_h, N_{hd}], \\
\texttt{gpu\_tensor* G'}
&\equiv
\mathrm{LayoutTransform}(\texttt{G}), \\
\mathrm{shape}(\texttt{G'})
&=
[ \texttt{N}_{\mathrm{npu\_channel}},
 -1,
 \texttt{tile\_chunk}].
\end{aligned}
\label{equation_transform_1}
\end{equation}
This arrangement increases the size of individual RDMA transfers and reduces the total number of RDMA requests.
\begin{figure}[!t]
\centering
\scalebox{0.98}{\includegraphics[width=1\columnwidth]{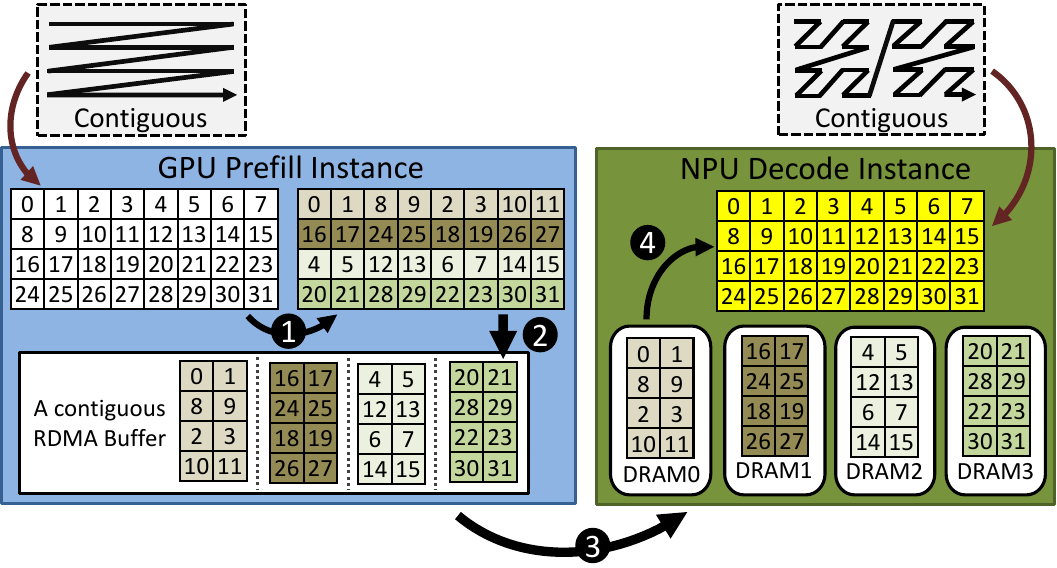}}
\vspace{-8pt}
\caption{
HAWK first partially aligns GPU tensors with the target NPU layout while forming large contiguous chunks for each per-channel NPU RDMA buffer. The remaining layout alignment is completed during transfer.
}
\label{rdma_transfer}
\vspace{-8pt}
\end{figure}
HAWK performs this transformation on the GPU prefill instance before transfer because prefill places less pressure on device memory bandwidth than decode, leaving sufficient bandwidth headroom for the additional memory operations introduced by the transformation.
The partially transformed tensor is then placed in a contiguous GPU RDMA buffer~(\ding{183}).
In the second step, HAWK pairs each contiguous portion of the partially transformed GPU tensor with the corresponding per-channel NPU RDMA buffer~(\ding{184}), completing the remaining layout alignment during transfer and producing the exact target NPU tensor layout without additional post-transfer transformation~(\ding{185}), as expressed in Eq~(\ref{equation_transform_2}). 
Here, $\texttt{T}_i$ denotes the NPU RDMA buffer for channel $i$,
$\Delta$ is the byte stride between channel regions in $\texttt{G}_B$,
and $\mathrm{size}_{\mathrm{transfer}}$ is the number of bytes
transferred to each buffer.
\begin{equation}
\begin{aligned}
\texttt{gpu\_tensor* G'}
&\xrightarrow{\mathrm{place}}
\texttt{gpu\_buffer* G}_B, \\
\texttt{npu\_buffer}\;\texttt{T}_{i}
&\equiv (ptr_{ch_i}, size_{ch_i}), \\
\bigl(\mathrm{ptr}(\texttt{G}_B) + i \cdot \Delta,
      size_{\mathrm{transfer}}\bigr)
&\xrightarrow{\mathrm{RDMA}}
\texttt{T}_{i}.
\end{aligned}
\label{equation_transform_2}
\end{equation}

One alternative is to map each small portion of the GPU tensor to the corresponding NPU buffer according to the target tensor layout, completing the layout transformation during transfer.
However, this approach significantly increases the number of RDMA requests and reduces the size of each transfer to only a few hundred bytes, increasing protocol and request-processing overhead and ultimately degrading transfer efficiency.
For example, in our setup, transferring the 20 MB KV cache for only 64 input tokens in 256 B portions requires approximately 80,000 RDMA requests.
In contrast, HAWK reduces transformation overhead by exploiting available memory-bandwidth headroom during prefill and decomposing the transformation into two steps, while improving transfer efficiency by aggregating data into fewer, larger RDMA transfers.

\begin{algorithm}[t]
\caption{End-to-End KV Cache Transfer with \textit{HAWK}}
\label{alg:rdma-transfer}
\begin{algorithmic}

\item[\textbf{Decode Instance: Request KV Cache}]
\item[D1:] sendKVRequest(\textit{prefillInstance}, \textit{requestId})

\item[]

\item[\textbf{Prefill Instance: Receive request and prepare for transfer}]
\item[P1:] \textit{requestId} $\gets$ recvKVRequest(\textit{decodeInstance})
\item[P2:] \textit{gpuKVTensor} $\gets$ getKVTensor(\textit{requestId})
\item[P3:] \textit{sendKVTensor} $\gets$ transformTensor(\textit{gpuKVTensor})
\item[P4:] \textit{sendBuffer} $\gets$ getRegisteredBuffer(size(\textit{sendKVTensor}))
\item[P5:] copyTensorToBuffer(\textit{sendBuffer}, \textit{sendKVTensor})
\item[P6:] \textit{sendBufferMetadata} $\gets$ getBufferMetadata(\textit{sendBuffer})
\item[P7:] sendMetadata(\textit{decodeInstance}, \textit{sendBufferMetadata})

\item[]

\item[\textbf{Decode Instance: Initiate Transfer \& store KV cache}]
\item[D2:] \textit{sendBufferMetadata} $\gets$ recvMetadata(\textit{prefillInstance})
\item[D3:] \textit{recvBufferList} $\gets$ getRegisteredBuffers(\textit{sendBufferMetadata})
\item[D4:] \textit{xferReqs} $\gets$ makeXferReqs(\textit{recvBufferList}, \textit{sendBufferMetadata})
\item[D5:] postBatchTransfer(\textit{xferReqs})
\item[D6:] waitForTransferCompletion(\textit{xferReqs})
\item[D7:] storeKVTensor(\textit{recvBufferList}, \textit{npuKVCachePool})
\item[]

\item[\textbf{Prefill \& Decode Instances: Wait for completion \& clean up}]
\item[D8:] notifyTransferCompletion(\textit{prefillInstance})
\item[D9:] releaseBuffers(\textit{recvBufferList})
\item[P8:] waitUntilNotified(\textit{decodeInstance})
\item[P9:] releaseBuffer(\textit{sendBuffer})

\end{algorithmic}
\end{algorithm}

\subsection{Putting It All Together}

Algorithm~\ref{alg:rdma-transfer} summarizes the end-to-end flow of KV cache transfer between GPU prefill and NPU decode instances, omitting the first-token transfer (a single integer) for simplicity.
First, the decode instance requests the KV cache associated with a request ID from the prefill instance (line~D1). The prefill instance then retrieves the corresponding KV cache, applies the transfer-aware transformation, reserves the required space in the GPU RDMA buffer, copies the tensor to the reserved region, and sends the associated metadata, including the total size and buffer offset, to the decode instance (lines~P1--P7). Using this metadata, the decode instance constructs RDMA requests by pairing regions of the GPU RDMA buffer with the corresponding per-channel NPU RDMA buffers (lines~D2--D5). These requests are submitted as a batch to avoid sequential RDMA request submission. After the transfer completes, the decode instance stores the received KV cache tensor in the NPU KV cache pool by copying
it from the NPU RDMA buffers (lines~D6--D7). Finally, both instances release the reserved regions in their RDMA buffers (lines~D8--D9 and P8--P9).

\subsection{Discussion}

\subsubsection{Extensibility Across NPU Architectures}

Diverse NPUs employ device-specific tensor layouts to optimize memory placement, data movement, and compute efficiency, and these layouts often differ from GPU tensor layouts~\cite{tt_tensor_layout, xla_tpu_layout, ascend_layout, qualcomm_tensor_layout}. Such layout differences can arise even when the same framework-level tensor is placed on different devices.
For example, TensorFlow tensors placed on Google TPUs~\cite{tensorflow_osdi, tpu_isca17, tpu_v4} use layouts that differ from those used on GPUs~\cite{xla_tpu_layout, xla_shapes_layout}, including differences in tiling, padding, and element ordering. These differences highlight the need for heterogeneity-aware data transfer across devices. In this context, HAWK provides a generalizable design for such transfers.
Our prototype implementation is configured for an optimized tiled tensor layout~\cite{tt_tensor_layout}, while many other NPUs similarly employ tiled or blocked tensor layouts~\cite{xla_tpu_layout,
ascend_layout}, allowing HAWK's core mechanisms to be adapted to these architectures. Even when tensor-layout details differ, HAWK can be adapted through device-specific changes to tensor-aware RDMA buffer allocation to accommodate the target NPU layout.

\subsubsection{Adaptability to Other Forms of Disaggregation}

Most data objects in LLM inference, including model weights, KV caches, and intermediate activations, are represented as tensors.
Since HAWK fundamentally targets efficient tensor transfer across heterogeneous devices, it can be adapted to other heterogeneous disaggregation schemes that require transferring tensor objects between devices.
For example, in operator-level disaggregation~\cite{pim_neupim, operator_mega2025sigcomm}, attention computation can be performed on GPUs while subsequent operators are executed on NPUs.
In such a configuration, intermediate activation tensors must be transferred between the two devices.
Compared with KV caches, intermediate activations are typically smaller, making fixed transformation and layout-adjustment overheads more significant relative to the transfer time. Moreover, their transfer lies between dependent operators, limiting opportunities for latency hiding through overlap.
In this context, HAWK can provide substantial benefits for such tensor transfers.

\subsubsection{Reverse Transfer}

While HAWK targets data transfer from GPUs to NPUs, the same approach can be extended to NPU-to-GPU transfers, which can be useful when an NPU offers higher prefill efficiency than a GPU.
In such cases, the transfer-aware transformation can be performed on the NPU, partially aligning the tensor with the target GPU layout while arranging portions destined for contiguous regions of the GPU buffer contiguously within each per-channel NPU buffer.
Each contiguous portion in the per-channel NPU RDMA buffers can then be mapped to the corresponding region of a virtually contiguous GPU RDMA buffer, completing the remaining layout alignment during transfer.
This preserves the benefit of fewer, larger RDMA transfers while allowing the transferred tensor to be used directly for GPU decoding without additional post-transfer layout transformations.

\section{Evaluation}

\subsection{Methodology}
\label{section_method}

\subsubsection{Hardware Configuration}
\label{section_hardware_config}

The system consists of a single prefill instance and a single decode instance: the prefill instance runs on NVIDIA A100 PCIe GPUs~\cite{a100}, interconnected within each node via NVLink Gen~3 and PCIe Gen~4, while the decode instance runs on Tenstorrent p150a boards~\cite{tt_p150a}, interconnected via Ethernet-based links.
We use Tenstorrent NPUs because, to the best of our knowledge, they are the only NPU platform that provides an open-source software stack permitting the low-level modifications required by HAWK.
The prefill and decode instances run on separate nodes connected via 200~Gbps HDR InfiniBand, with each node equipped with a ConnectX-6 NIC.

\subsubsection{Software Configuration}
\label{section_software_config}

We use vLLM v0.10.0~\cite{vllm} across all evaluated systems to provide a common serving framework for comparison and demonstrate HAWK's integration into a widely used LLM serving environment. To support NPU execution, we integrate the Tenstorrent LLM stack with vLLM through a plugin. This integration may introduce overhead that reduces NPU serving performance relative to standalone execution.
We retain the default mixed-precision configuration provided by the Tenstorrent LLM software stack~\cite{tt_metal}.
To assess output quality under this configuration, we use ROUGE, a standard metric for summarization quality~\cite{acc-lin-2004-rouge,acc_see-etal-2017-get}, to compare the two configurations on the evaluated workloads.
Relative to the homogeneous GPU-only setup, the heterogeneous GPU-NPU setup achieves 0.74\% higher ROUGE-1, 2.28\% lower ROUGE-2, and 3.12\% higher ROUGE-L scores.

For GPU-side RDMA support, we use \texttt{nvidia-peermem}~\cite{cuda_rdma_document, open_driver_nvidia}. On the NPU side, we extend the Tenstorrent software stack (TT-Metal v0.68.0~\cite{tt_metal}, the user-mode driver~\cite{tt_umd}, and the kernel-mode driver v2.5.0~\cite{tt_kmd}) to implement the RDMA-related functionality required by HAWK, including a peer memory client and tensor-layout-aware RDMA buffer allocation.
Building on this support, we implement a user-space transfer mechanism on top of Mooncake Transfer Engine v0.3.6~\cite{mooncake,mooncake_github}, a high-performance data transfer engine for disaggregated LLM serving. Our mechanism integrates transfer-aware tensor transformation with RDMA buffer pairing.
For the model, we use Llama 3.1 70B~\cite{model_llama32024herdmodels}, whose scale requires multiple accelerators and is therefore suitable for evaluating phase-disaggregated serving. We distribute the model across four devices per instance using tensor parallelism (TP). In practice, the current Tenstorrent LLM stack tightly couples each supported model to a specific device type and device count, which limits our evaluation to this model~\cite{tt_model}.

\begin{table}[!t]
\renewcommand{\arraystretch}{1.00}
  \centering
  \resizebox{1.00\columnwidth}{!}{
  \begin{tabular}{R{2.7cm}|L{5.3cm}}
    \hline
    \cline{1-2}
    \tablefont{\textbf{Component}} & \tablefont{\textbf{Configuration}} \\
    \hline
    \cline{1-2}
    \tablesmall{Prefill hardware} & \tablesmall{4$\times$ NVIDIA A100 PCIe 40~GB GPUs}  \\
    \hline
    \tablesmall{GPU decode hardware} & \tablesmall{4$\times$ NVIDIA A100 PCIe 80~GB GPUs}  \\
    \hline
    \tablesmall{NPU decode hardware} & \tablesmall{4$\times$ Tenstorrent p150a NPUs}  \\
    \hline
    \tablesmall{Inter-node network} & \tablesmall{200~Gbps HDR InfiniBand, \newline ConnectX-6 NIC per node} \\
    \hline
    \tablesmall{Serving setup} & \tablesmall{vLLM, one instance per phase, \newline TP=4 per instance} \\
    \hline
    \tablesmall{Model} & \tablesmall{Llama 3.1 70B (148~GB model weights)} \\
    \hline
    \cline{1-2}
  \end{tabular}
  }
\end{table}

\begin{table}[!t]
\renewcommand{\arraystretch}{1.00}
  \centering
  \resizebox{1.00\columnwidth}{!}{
  \begin{tabular}{R{1.8cm}|C{1.8cm}|C{1.0cm}C{1.0cm}|C{1.0cm}C{1.0cm}}
    \hline
    \cline{1-6}
    \multirow{2}{*}{\tablefont{\textbf{Workloads}}} & \multirow{2}{*}{\tablefont{\textbf{Dataset}}} & \multicolumn{2}{c|}{\tablefont{\textbf{Input Length}}} & \multicolumn{2}{c}{\tablefont{\textbf{Output Length}}} \\
    \cline{3-6}
    \tablefont{} & \tablefont{} & \tablefont{Mean} & \tablefont{P90} & \tablefont{Mean} & \tablefont{P90} \\
    \hline
    \cline{1-6}
    \tablesmall{Chatbot} & \tablesmall{ShareGPT~\cite{sharegpt_vicuna_unfiltered}} & \tablesmall{783} & \tablesmall{899} & \tablesmall{140} & \tablesmall{148} \\
    \hline
    \tablesmall{Summarization} & \tablesmall{LongBench~\cite{bai2024longbench}} & \tablesmall{2186} & \tablesmall{2860} & \tablesmall{86} & \tablesmall{115}\\
    \hline
    \cline{1-6}
  \end{tabular}
  }
  \vspace{3pt}
\caption{
System and online serving workload configurations.
}
  \vspace{-20pt}
  \label{table:system_config}
\end{table}

\subsubsection{Comparison Configurations}
We evaluate four configurations: three baselines and HAWK.

\noindent\textbf{1. GPU-GPU + Host-Staging (GG-Host)} uses GPUs for both prefill and decode and transfers the KV cache through host staging.

\noindent\textbf{2. GPU-NPU + Host-Staging (GN-Host)} uses GPUs for prefill and NPUs for decode and transfers the KV cache through host staging, followed by tensor-layout adjustment on the NPU side.

\noindent\textbf{3. GPU-GPU + Direct RDMA (GG-Direct)} uses GPUs for both prefill and decode and transfers the KV cache directly between GPU memories via RDMA using the Mooncake Transfer Engine. We verify that its RDMA transfer performance is comparable to that of other GPU RDMA transfer engines, such as NIXL~\cite{nvidia_nixl}.

\noindent\textbf{4. GPU-NPU + HAWK (GN-HAWK)} uses GPUs for prefill and NPUs for decode and transfers the KV cache using HAWK.

\vspace{2pt}
One additional configuration uses GPUs for prefill and NPUs for decode, transferring the KV cache directly between device memories via RDMA while using a conventional single contiguous RDMA buffer on the NPU side. This setup incurs post-transfer adjustment overhead because the buffer does not conform to the NPU tensor layout.
We implement a prototype of this configuration, but it fails with an out-of-memory error during serving due to NPU memory fragmentation: the buffer creates imbalanced memory usage across DRAM channels, leaving insufficient capacity for subsequent tensor allocations. 
In our implementation, allocating a $Y$~MB buffer in one of the p150a NPU's eight DRAM channels leaves $7Y$~MB of memory underutilized across the remaining seven channels due to this imbalance.
We therefore exclude this configuration from the quantitative comparison.
This failure highlights the importance of HAWK's tensor-layout-aware RDMA buffer allocation in preventing memory fragmentation, which persists even when tensor transformation overhead becomes
negligible relative to inter-node transfer time or is partially hidden by overlapping transformation with computation.

\subsubsection{Workloads}

We evaluate our system using two types of workloads. First, we use a synthetic workload to evaluate KV cache transfer overhead. Second, we evaluate end-to-end LLM serving performance using workloads derived from prior work on phase-disaggregated LLM serving~\cite{distserve}.
Each serving workload is derived from a different dataset~\cite{sharegpt_vicuna_unfiltered, bai2024longbench} and exhibits different input/output token lengths and request sizes, resulting in different KV cache transfer characteristics.
For each workload, we vary the average request arrival rate and assume a Poisson arrival process.

\vspace{2pt}
Table~\ref{table:system_config} summarizes the evaluation setup described above.
We plan to release our implementation as open source.

\subsection{End-to-End KV Cache Transfer Latency}
\label{section_eval}

\begin{figure}[!t]
\centering
\scalebox{1.00}{\includegraphics[width=1.0\columnwidth]{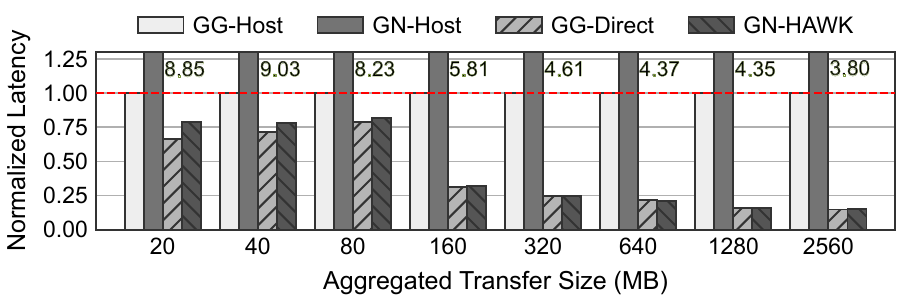}}
\vspace{-20pt}
\caption{
End-to-end KV cache transfer latency across aggregated transfer sizes, normalized to GG-Host.
GN-HAWK achieves latency comparable to GG-Direct, differing by only 0.39\% at 2560~MB.
}
\label{compare_transfer}
\vspace{-12pt}
\end{figure}

Figure~\ref{compare_transfer} compares end-to-end KV cache transfer latency across aggregated transfer sizes with a tensor parallelism degree of four. Latency is measured from when the decode instance requests the KV cache to when the received cache is stored in its KV cache pool. As the aggregated transfer size increases from 20 MB to 2560 MB, GN-Host incurs 8.85$\times$ to 3.80$\times$ the latency of GG-Host. Although the relative gap narrows as heterogeneity-related overheads are amortized over longer data-transfer times, GN-Host remains substantially slower. This result shows the importance of heterogeneity-aware transfer support for efficient cross-device data transfer.

Compared with host-staged transfers, direct transfers exhibit substantially lower latency. Over the same size range, GG-Direct's normalized latency decreases from 66.42\% to 14.86\% of GG-Host latency, while GN-HAWK's decreases from 79.08\% to 14.92\%. GN-HAWK incurs slightly higher latency than GG-Direct for small transfers because it additionally transforms tensors and constructs multiple RDMA requests for the per-channel NPU RDMA buffers. These overheads arise from differences in tensor layout and memory placement across devices, but their relative impact decreases at larger transfer sizes.

\begin{figure}[!t]
\begin{subfigure}{0.491\columnwidth}
    \captionsetup{skip=0pt}
    \includegraphics[width=\linewidth]{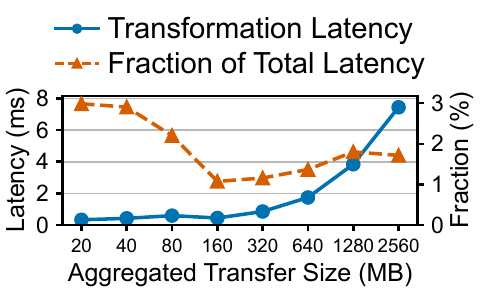}
    \caption{Transformation latency}
    \label{break_1}
\end{subfigure}
\begin{subfigure}{0.491\columnwidth}
    \captionsetup{skip=0pt}
    \includegraphics[width=\linewidth]{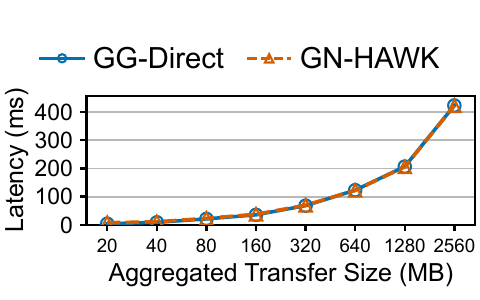}
    \caption{Transfer-pipeline latency}
    \label{break_2}
\end{subfigure}
\caption{
KV cache transfer overhead: (a) tensor-transformation latency and its fraction of end-to-end latency, and (b) transfer-pipeline latency.
At 2560~MB, transformation accounts for 1.71\% of GN-HAWK's end-to-end latency, while transfer-pipeline latency differs from GG-Direct by only 0.25\%.
}
\label{eval_breakdown}
\end{figure}

To examine these overheads, Figure~\ref{eval_breakdown} shows the latencies of tensor transformation and the transfer pipeline separately.
Figure~\ref{break_1} shows that transformation latency increases from 0.33 ms at 20 MB to 7.44 ms at 2560 MB, while its share of GN-HAWK's end-to-end latency decreases from 2.98\% to 1.71\%. In GN-Host, layout conversion is performed within the NPU's host-to-device transfer API, coupling conversion with costly host-staged data movement. HAWK instead partially transforms tensors on the GPU and completes layout alignment during direct RDMA transfer, avoiding this host-staged path.
Figure~\ref{break_2} shows transfer-pipeline latency, including request mapping, inter-node data transfer, and synchronization. Although GN-HAWK incurs relatively higher latency for small transfers, its transfer-pipeline latency differs from GG-Direct by only 0.25\% at 2560 MB. At the same size, the end-to-end latency difference is only 0.39\%, showing that HAWK limits the additional overhead of heterogeneous transfer.
In our setup, a 20 MB KV cache corresponds to only 64 input tokens, representing a short input sequence. These results show that the relative impact of HAWK's additional transformation and request-processing overhead decreases as transfer size increases, enabling end-to-end latency within 1\% of GPU-to-GPU RDMA at 2560 MB.

\subsection{Online Serving Performance}

For online serving, we evaluate four metrics: (1) time to first token (TTFT), (2) time per output token (TPOT), (3) service-level objective (SLO) attainment, and (4) output throughput. These metrics can be categorized into two groups based on their sensitivity to KV cache transfer. Transfer-sensitive metrics include TTFT, SLO attainment, and decode throughput, whereas TPOT is largely insensitive to KV cache transfer and instead depends primarily on the underlying decode execution characteristics.
We set the TTFT/TPOT SLOs to \textit{2000/200 ms} for Chatbot and \textit{15000/200 ms} for Summarization, following MLPerf Inference~\cite{slo_llama} and prior work~\cite{distserve}.
TTFT and TPOT SLO attainment measure the fractions of requests meeting their respective targets. Request SLO attainment measures the fraction of requests meeting both the TTFT and TPOT targets.

\begin{figure}[!t]
\begin{subfigure}{0.95\columnwidth}
    \includegraphics[width=\linewidth]{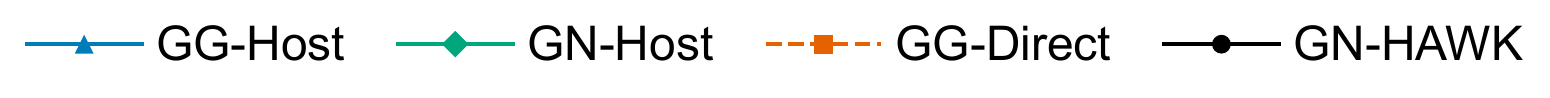}
\end{subfigure}
\vspace{0pt}

\begin{subfigure}{0.490\columnwidth}
    \captionsetup{skip=0pt}
    \includegraphics[width=\linewidth]{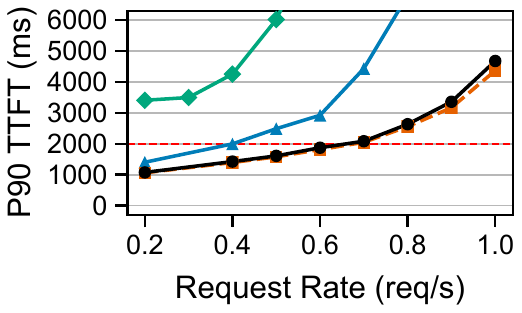}
\caption{Chatbot.}
    \label{ttft_1}
\end{subfigure}
\hspace{0pt}
\begin{subfigure}{0.490\columnwidth}
    \captionsetup{skip=0pt}
    \includegraphics[width=\linewidth]{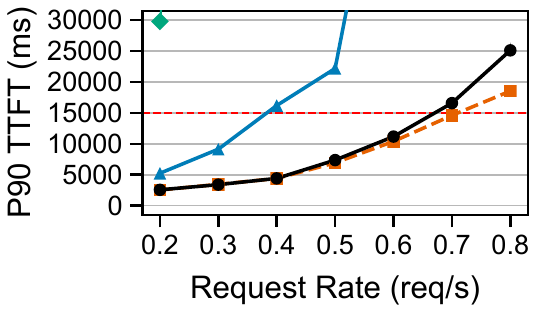}
\caption{Summarization.}
    \label{ttft_2}
\end{subfigure}

\vspace{-5pt}
\caption{
P90 TTFT under varying request rates. At 90\% TTFT-SLO attainment, GN-HAWK sustains request rates within 2.90\% and 5.90\% of GG-Direct for Chatbot and Summarization.
}
\vspace{-10pt}
\label{eval_ttft}
\end{figure}

\subsubsection{TTFT}

Figure~\ref{eval_ttft} shows P90 TTFT across configurations together with the corresponding SLO thresholds.
GG-Host incurs higher TTFT than GG-Direct and GN-HAWK due to longer KV cache transfer latency, reaching the SLO limits at 0.40 and 0.38 requests/s for Chatbot and Summarization, respectively. Additional heterogeneous transfer overhead further increases TTFT for GN-Host, making it difficult to meet the SLO even at low request rates.
In contrast, GN-HAWK closely tracks GG-Direct, with request-rate differences of only 2.90\% and 5.90\% at 90\% SLO attainment for the two workloads, respectively, demonstrating HAWK's effectiveness in mitigating heterogeneous transfer overhead.
For Summarization, slower NPU decode execution increases backpressure and request queuing at the prefill node, widening the TTFT gap at higher request rates. Further optimization of NPU serving performance could help narrow this gap.

\begin{figure}[!t]
\begin{subfigure}{0.95\columnwidth}
    \includegraphics[width=\linewidth]{figures/5_eval/legend.pdf}
\end{subfigure}
\vspace{0pt}

\begin{subfigure}{0.490\columnwidth}
    \captionsetup{skip=0pt}
    \includegraphics[width=\linewidth]{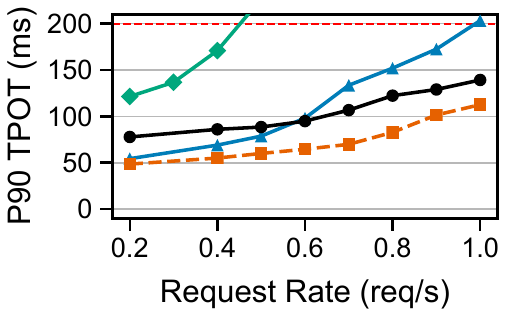}
\caption{Chatbot.}
    \label{tpot_1}
\end{subfigure}
\hspace{0pt}
\begin{subfigure}{0.490\columnwidth}
    \captionsetup{skip=0pt}
    \includegraphics[width=\linewidth]{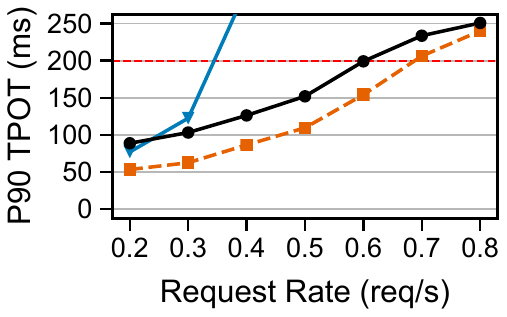}
\caption{Summarization.}
    \label{tpot_2}
\end{subfigure}

\vspace{-5pt}
\caption{
P90 TPOT under varying request rates. GN-HAWK exhibits a small TPOT gap with GG-Direct due to differences in decode performance. However, GN-HAWK meets the SLO for Chatbot, and the gap narrows at higher request rates for Summarization.
}
\vspace{-10pt}
\label{eval_tpot}
\end{figure}

\subsubsection{TPOT}

Figure~\ref{eval_tpot} shows P90 TPOT across configurations together with the corresponding SLO thresholds. For Chatbot, the TPOT of both GG-Direct and GN-HAWK remains below the SLO threshold. For Summarization, GN-HAWK reaches the TPOT SLO threshold at a lower request rate (i.e., 0.60 requests/s) than GG-Direct due to its slightly lower decode performance. However, the TPOT gap narrows substantially at a higher request rate (i.e., 0.80 requests/s). At higher request rates, requests are more likely to be processed in larger batches, and the NPU in our evaluation achieves higher decode efficiency at larger batch sizes, bringing its performance closer to that of the GPU. The results show that NPUs can achieve competitive serving performance, although small performance gaps remain. These gaps may be further reduced by optimizing system configurations, such as batch size, or by using other NPU platforms.
For the host-staged configurations, GG-Host and GN-Host exhibit higher TPOT for the two workloads, which we attribute to host intervention in the evaluated vLLM implementation interfering with decode execution.

\begin{figure}[!t]
\begin{subfigure}{0.95\columnwidth}
    \includegraphics[width=\linewidth]{figures/5_eval/legend.pdf}
\end{subfigure}
\vspace{0pt}

\begin{subfigure}{0.490\columnwidth}
    \captionsetup{skip=0pt}
    \includegraphics[width=\linewidth]{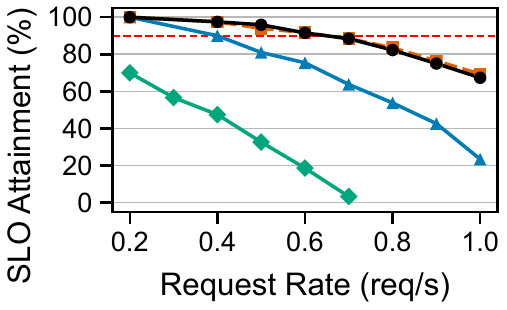}
\caption{Chatbot.}
    \label{slo_1}
\end{subfigure}
\hspace{0pt}
\begin{subfigure}{0.490\columnwidth}
    \captionsetup{skip=0pt}
    \includegraphics[width=\linewidth]{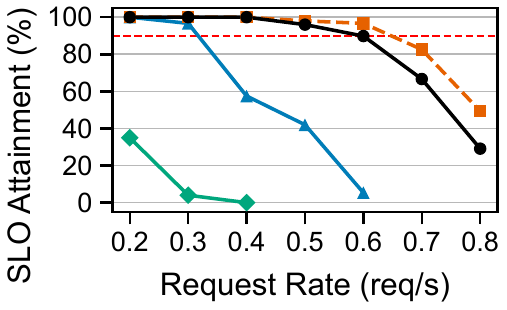}
\caption{Summarization.}
    \label{slo_2}
\end{subfigure}

\vspace{-5pt}
\caption{
SLO attainment under varying request rates.
GN-HAWK closely follows GG-Direct for Chatbot.
Despite showing slightly lower attainment for Summarization, the difference in request rate at 90\% SLO attainment is only 7.69\% relative to GG-Direct.
}
\vspace{-10pt}
\label{eval_slo}
\end{figure}

\subsubsection{SLO Attainment}

Figure~\ref{eval_slo} shows request SLO attainment across configurations, reflecting the combined impact of TTFT and TPOT.
For Chatbot, GG-Host falls below 90\% SLO attainment at 0.40 requests/s, primarily due to the TTFT violations.
GN-Host performs even worse, failing to achieve 90\% SLO attainment even at low request rates because of severe TTFT violations.
In contrast, GG-Direct and GN-HAWK maintain 90\% SLO attainment up to 0.65 requests/s, benefiting from their lower transfer overhead and TTFT.
In particular, GN-HAWK closely tracks GG-Direct by effectively mitigating heterogeneous transfer overhead.
For Summarization, GG-Host and GN-Host similarly exhibit poor SLO attainment, whereas GG-Direct and GN-HAWK sustain substantially higher attainment.
GN-HAWK reaches the 90\% SLO threshold at 0.60 requests/s, slightly earlier than GG-Direct at 0.65 requests/s.
This difference is primarily attributable to TPOT violations caused by the slightly lower NPU decode performance.
Nevertheless, the request-rate difference at 90\% SLO attainment is only 7.69\%, indicating that GN-HAWK achieves performance comparable to GG-Direct.

\begin{figure}[!t]
\vspace{-2pt}
\hspace{2pt}
\begin{subfigure}{0.75\columnwidth}
    \includegraphics[width=\linewidth]{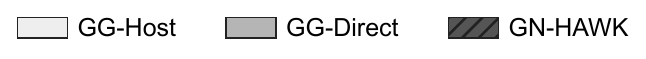}
\end{subfigure}
\vspace{-5pt}

\begin{subfigure}{0.490\columnwidth}
    \captionsetup{skip=0pt}
    \includegraphics[width=\linewidth]{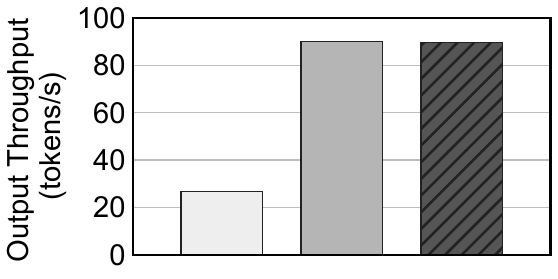}
\caption{Chatbot.}
\label{decode_1}
\end{subfigure}
\hspace{0pt}
\begin{subfigure}{0.490\columnwidth}
    \captionsetup{skip=0pt}
    \includegraphics[width=\linewidth]{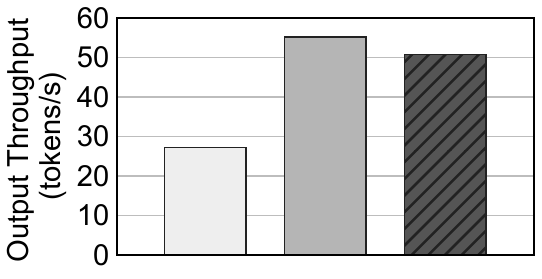}
\caption{Summarization.}
\label{decode_2}
\end{subfigure}

\vspace{-5pt}
\caption{
Output throughput at 90\% SLO attainment across configurations. GN-Host is omitted because it fails to achieve 90\% SLO attainment even at the lowest evaluated request rate.
GN-HAWK shows a decode throughput gap of only up to 7.88\% compared to GG-Direct.
}
\vspace{-5pt}
\label{eval_decode}
\end{figure}

\subsubsection{Output Throughput}

Figure~\ref{eval_decode} shows the output throughput of each configuration at the point where it reaches 90\% SLO attainment.
We exclude GN-Host because it fails to achieve 90\% SLO attainment even at the lowest evaluated request rate.
GG-Host shows lower decode throughput for both workloads than GG-Direct and GN-HAWK due to its transfer overhead.
GN-HAWK achieves only 0.48\% and 7.88\% lower decode throughput than GG-Direct for Chatbot and Summarization, respectively.
The remaining small throughput gap mainly stems from differences in decode device performance, while HAWK limits additional overhead from heterogeneous integration.
These results show that GPU-NPU heterogeneous integration with proper data transfer support can achieve performance comparable to that of a GPU-only homogeneous system under serving SLO constraints.

\subsection{Serving Efficiency}
\label{section_eff}

The main benefit of GPU-NPU heterogeneous integration with HAWK lies in power and cost efficiency.
We compare these efficiencies at 90\% SLO attainment using output token throughput, focusing on GG-Direct and GN-HAWK as the best-performing homogeneous and heterogeneous configurations, respectively.

\subsubsection{Power Efficiency}
\label{section_power}
We calculate power efficiency by dividing output token throughput by the aggregate ASIC power of all prefill and decode devices, averaged over the same measurement interval.
Specifically, we use device ASIC power, including compute cores and on-chip SRAM, because the available Tenstorrent monitoring interface reports only NPU core power~\cite{tt_smi}.
To estimate GPU ASIC power, we first collect board-level power measurements using NVIDIA DCGM~\cite{nvidia_dcgm}.
We then collect binary traces of GPU kernels executed during prefill and decoding using NVBit~\cite{nvbit} and feed them into a detailed GPU power simulator to estimate the fraction of total power attributable to the GPU ASIC for each phase~\cite{accelwatch}.
Finally, we apply these phase-specific fractions to the corresponding board-level power measurements.
Regarding excluded power components, both configurations use the same interconnect hardware and exhibit similar transfer times, suggesting that differences in interconnect energy consumption may be limited.
We also consider memory power. Although HBM2e in the evaluated GPU offers higher bandwidth per watt than GDDR6 in the evaluated NPU, estimates accounting for their respective memory bandwidths indicate higher memory power consumption for the GPU~\cite{micron_power}.

As shown in Table~\ref{table:efficiency}, GN-HAWK achieves 1.26$\times$ and 1.15$\times$ higher power efficiency than GG-Direct for Chatbot and Summarization, respectively.
With an assumed 5\% error in GPU power measurements, these gains remain at least 1.23$\times$ and 1.13$\times$, respectively.
These gains come from combining GPU prefill with NPU decode to improve efficiency while maintaining comparable serving performance.
HAWK makes this combination effective by reducing transfer overhead between the heterogeneous devices.
Although the magnitude of the gains may vary across NPU platforms, the results demonstrate the potential of GPU-NPU heterogeneous serving to improve power efficiency.

\begin{table}[!t]
\renewcommand{\arraystretch}{0.90}
  \centering
  \resizebox{1.0\columnwidth}{!}{
  \begin{tabular}{C{2.9cm}|L{1.4cm}|C{2.0cm}C{2.0cm}}
    \hline
    \cline{1-4}
    \multicolumn{2}{c|}{\multirow{2}{*}{\tablefont{\textbf{Performance}}}}
    & \multicolumn{2}{c}{\tablesmall{\textbf{Workloads}}} \\
    \cline{3-4}
    \multicolumn{2}{c|}{}
    & \tablesmall{Chatbot}
    & \tablesmall{Summarization} \\
    \hline
    \cline{1-4}
    \tablesmall{Serving power efficiency} &\tablesmall{GG-Direct}
    & \tablefont{67.09} & \tablefont{41.01}  \\
    \tablesmall{(output tokens/s/kW)} & \tablesmall{GN-HAWK} 
    & \tablefont{\textbf{84.48}} & \tablefont{\textbf{47.18}} \\
    \hline
    \tablesmall{Serving cost efficiency} &\tablesmall{GG-Direct}
    & \tablefont{2.05} & \tablefont{1.25} \\
    \tablesmall{(output tokens/s/\$1,000)} & \tablesmall{GN-HAWK} 
    & \tablefont{\textbf{4.15}} & \tablefont{\textbf{2.35}} \\
    \hline
    \cline{1-4}
  \end{tabular}
  }
  \vspace{3pt}
\caption{
Serving power and cost efficiency comparison between GG-Direct and GN-HAWK at 90\% SLO attainment.
GN-HAWK achieves up to 1.26$\times$ higher power efficiency and 2.03$\times$ higher cost efficiency.
}
  \vspace{-19pt}
  \label{table:efficiency}
\end{table}

\begin{table}[!t]
\renewcommand{\arraystretch}{0.90}
  \centering
  \resizebox{1.0\columnwidth}{!}{
  \begin{tabular}{C{2.9cm}|L{1.4cm}|C{2.0cm}C{2.0cm}}
    \hline
    \cline{1-4}
    \multicolumn{2}{c|}{\multirow{2}{*}{\tablefont{\textbf{Performance}}}}
    & \multicolumn{2}{c}{\tablesmall{\textbf{Workloads}}} \\
    \cline{3-4}
    \multicolumn{2}{c|}{}
    & \tablesmall{Chatbot}
    & \tablesmall{Summarization} \\
    \hline
    \cline{1-4}
    \tablesmall{Serving throughput} &\tablesmall{GGH-Direct}
    & \tablefont{95.96} & \tablefont{54.28}  \\
    \tablesmall{(output tokens/s)} & \tablesmall{GN-HAWK}
    & \tablefont{\textbf{89.69}} & \tablefont{\textbf{50.75}} \\
    \hline
    \tablesmall{Serving power efficiency} &\tablesmall{GGH-Direct}
    & \tablefont{81.53} & \tablefont{45.31}  \\
    \tablesmall{(output tokens/s/kW)} & \tablesmall{GN-HAWK} 
    & \tablefont{\textbf{84.48}} & \tablefont{\textbf{47.18}} \\
    \hline
    \tablesmall{Serving cost efficiency} &\tablesmall{GGH-Direct}
    & \tablefont{1.20} & \tablefont{0.68} \\
    \tablesmall{(output tokens/s/\$1,000)} & \tablesmall{GN-HAWK}
    & \tablefont{\textbf{4.15}} & \tablefont{\textbf{2.35}} \\
    \hline
    \cline{1-4}
  \end{tabular}
  }
  \vspace{3pt}
\caption{
Serving throughput, power efficiency, and cost efficiency comparison between GGH-Direct (GG-Direct with H100 GPUs for decoding) and GN-HAWK at 90\% SLO attainment.
GN-HAWK achieves 1.04$\times$ higher power efficiency and 3.46$\times$ higher cost efficiency than GGH-Direct, while maintaining a throughput difference within 6.53\%.
}
  \vspace{-12pt}
  \label{table:h100}
\end{table}

\subsubsection{Cost Efficiency}
\label{section_cost}

Since device prices vary with market conditions, we report cost efficiency as a reference rather than a definitive comparison across configurations.
Cost efficiency is calculated by dividing output token throughput by the aggregate cost of all prefill and decode devices.
We use the official vendor price for the NPU~\cite{tt_p150a} and publicly listed market prices for the GPUs~\cite{jarvislabs_a100_price, computeexchange_a100_price}: \$1,400 per p150a, \$4,000 per A100 40GB, \$7,000 per A100 80GB, and \$16,000 per H100.
As shown in Table~\ref{table:efficiency}, GN-HAWK achieves 2.03$\times$ and 1.88$\times$ higher cost efficiency than GG-Direct for Chatbot and Summarization, respectively.
These results suggest that GPU-NPU heterogeneous systems can be an economically attractive option for LLM serving with proper system support.

\subsubsection{Comparison with Another GPU Architecture}
\label{section_h100}

To examine whether the efficiency benefits of our proposed system hold with a different GPU architecture, we replace the decode-side A100 GPUs in GG-Direct with NVIDIA H100 GPUs~\cite{nvidia_h100_whitepaper}, refer to this configuration as \textit{GGH-Direct}, and compare it with GN-HAWK.
As shown in Table~\ref{table:h100}, GN-HAWK achieves only 6.53\% and 6.50\% lower output token throughput than GGH-Direct at 90\% SLO attainment for Chatbot and Summarization, respectively.
In terms of power efficiency, GN-HAWK achieves 1.04$\times$ higher efficiency than GGH-Direct for both workloads.
With an assumed 5\% error in GPU power measurements, these gains remain at least 1.02$\times$ for both workloads.
These gains are smaller than in the A100 comparison because the H100 GPUs consume less power during decoding in our evaluation.
For cost efficiency, GN-HAWK achieves 3.46$\times$ higher efficiency for both Chatbot and Summarization, reflecting the higher H100 prices used in our evaluation~\cite{computeexchange_h100_price}.
These results further suggest the potential of GPU-NPU heterogeneous serving to improve efficiency, with the gains depending on the specific GPU and NPU platforms used.

\section{Related Work}

\subsection{Phase-Disaggregation}
Prior work has proposed diverse optimizations for prefill-decode disaggregated serving systems, including phase-aware resource provisioning and placement~\cite{splitwise_isca, distserve, dp_hu2024inference}, heterogeneous GPU deployment~\cite{dp_jiang2025hexgen, dp_vendor_chen2025disaggregated}, cluster-level scheduling and dynamic prefill-decode ratio adaptation~\cite{dp_pdserve, mooncake, dp_ruan2025dynaserve, dp_wu2025arrow, dp_wang2025prefill}, and KV cache management and transfer optimization~\cite{mooncake, dp_hu2024memserve, dp_hong2025semi}.
These efforts primarily optimize disaggregated serving within GPU-based systems. We extend this scope to GPU-NPU phase disaggregation and demonstrate its potential for LLM serving.

\subsection{Other Forms of Heterogeneous Integration}
Prior work has explored LLM inference systems that combine GPUs with CPUs~\cite{add_lin2026hybridgen,cpu_gpu_he2024fastdecode,cpu_gpu_jiang2025neo,cpu_gpu_kamahori2025fiddler}, NPUs~\cite{hybe_npu,gpu_npu_mobil_soc_sosp,gpu_npu_wei2025agent}, or PIM/PNM accelerators~\cite{attacc_pim,pim_pact_near,pim_paise,aide}, as well as integrated NPU-PIM systems~\cite{pim_IANUS,pim_stanford_chen2025p3,pim_neupim}.
These studies exploit hardware heterogeneity through workload partitioning, offloading, scheduling, and disaggregation to improve inference performance and efficiency. Against this backdrop, we investigate the practical challenges of coordinating GPUs and NPUs across multiple servers and develop system-level solutions to address them.

\section{Conclusion}

We propose a GPU–NPU phase-disaggregated LLM inference system that leverages GPUs for prefill and NPUs for decode to achieve both high performance and efficiency. Realizing such a system, however, requires an efficient data-transfer path that addresses heterogeneity between GPUs and NPUs. To this end, we present HAWK, which enables direct data transfer over a common RDMA stack while addressing tensor-layout discrepancies through tensor-layout-aware RDMA buffer allocation and transfer-aware tensor transformation.
Our evaluation shows that the heterogeneous system with HAWK achieves performance within 7.88\% of the GPU-only system under the same SLO constraints, with up to 1.26$\times$ higher power efficiency and 2.03$\times$ higher cost efficiency.
\section*{Acknowledgments}

ChatGPT and Codex were used to assist with text editing,
grammar correction, LaTeX formatting, and developing scripts
to visualize experimental data.
Claude was used to assist with portions of the integration
of the Tenstorrent software stack into vLLM.
Generative AI was not used to develop the main idea or the code implementing the mechanisms proposed in HAWK.

\bibliographystyle{ACM-Reference-Format}
\bibliography{sample-base}

\end{document}